\documentclass[useAMS,usegraphicx,usenatbib]{mn2e}
\usepackage{amssymb}
\usepackage{times}
\newcommand{\D}[2]{\frac{\partial #2}{\partial #1}}

\newcommand\bb[1] {\mbox{\boldmath{$#1$}}}
\newcommand\del{\bb{\nabla}}
\newcommand\bcdot{\,\bb{\cdot}\,}
\newcommand\btimes{\,\bb{\times}\,}
\newcommand{\mc}[1]{\mathcal{#1}}
\newcommand{\ey}{\hat{\bb{e}}_y}
\newcommand{\ez}{\hat{\bb{e}}_z}

\newcounter{eqold}
\newcounter{eqbid}

\title[Interstellar Cloud Formation]
{Formation of Interstellar Clouds: Parker Instability with Phase Transitions}
\author[Mouschovias, Kunz \& Christie]
{Telemachos Ch. Mouschovias, Matthew W. Kunz \& Duncan A. Christie\\
Departments of Physics and Astronomy, University of Illinois at
Urbana-Champaign, 1002 W. Green Street, Urbana, IL  61801}
\date{Accepted 2009 January 07. Received 2009 January 06; in original form 2008 October 08}
\pagerange{\pageref{firstpage}--\pageref{lastpage}} \pubyear{2009}
\def\LaTeX{L\kern-.36em\raise.3ex\hbox{a}\kern-.15em
    T\kern-.1667em\lower.7ex\hbox{E}\kern-.125emX}
\begin{document}
\label{firstpage} \maketitle
\begin{abstract}
We follow numerically the nonlinear evolution of the Parker
instability in the presence of phase transitions from a warm to a
cold H\textsc{i} interstellar medium in two spatial dimensions.
The nonlinear evolution of the system favors modes that allow the
magnetic field lines to cross the galactic plane. Cold H\textsc{i}
clouds form with typical masses $\simeq 10^5$ M$_\odot$, mean
densities $\simeq 20$ cm$^{-3}$, mean magnetic field strengths
$\simeq 4.3\,\mu$G (rms field strengths $\simeq 6.4\,\mu$G),
mass-to-flux ratios $\simeq 0.1-0.3$ relative to critical,
temperatures $\simeq 50$ K, (two-dimensional) turbulent
velocity dispersions $\simeq 1.6$ km s$^{-1}$, and separations
$\simeq 500$ pc, in agreement with observations. The maximum density
and magnetic field strength are $\simeq 10^3$ cm$^{-3}$ and $\simeq
20$ $\mu$G, respectively. Approximately 60\% of all H\textsc{i}
mass is in the warm neutral medium. The cold neutral medium is
arranged into sheet-like structures both perpendicular and
parallel to the galactic plane, but it is also found almost
everywhere in the galactic plane, with the density being highest
in valleys of the magnetic field lines. `Cloudlets' also form
whose physical properties are in quantitative agreement with those
observed for such objects by Heiles (1967). The nonlinear phase of
the evolution takes $\lesssim 30$ Myr, so that, if the instability
is triggered by a nonlinear perturbation such as a spiral density
shock wave, interstellar clouds can form within a time suggested
by observations.
\end{abstract}

\begin{keywords}
instabilities --- ISM: clouds --- ISM: magnetic fields --- ISM: structure --- MHD
\end{keywords}

\section{Introduction}\label{S_intro}

Understanding the formation of interstellar clouds and cloud
complexes is a problem of longstanding importance, one which has
received increased attention in recent years. Since H\textsc{i}
clouds are the progenitors of molecular clouds, the stellar
birthplaces, the process by which they form is of fundamental
importance to galactic evolution as well as to the early stages of
star formation.

Proposed mechanisms for interstellar cloud formation fall into
three categories: stochastic coagulation of smaller clouds
\citep{oort54,fs65,kwan79,sh79}, colliding streams of turbulent
flows \citep{bphvs99,hbpb01,hsdhb06}, or collective effects involving
instabilities \citep{mouschovias74,msw74,bs80}. Stochastic
coagulation is not feasible because collisional agglomeration
proceeds too slowly and, also, collisions are more likely to
result in disruption rather than merger. Turbulence-driven formation
of atomic-hydrogen clouds is problematic because no realistic
mechanism has yet been demonstrated to sustain the required
large-scale, ordered, focused motions for $\simeq 10$ Myr. In
addition, the short lifetimes and highly supercritical mass-to-flux
ratios of molecular clouds thought to form within these turbulent
`atomic precursors' are in conflict with observations \citep{mtk06}.
\footnote{The typical separation between the {\it
first} appearance of CO and the {\em peak} of H$\alpha$ in spiral
arms corresponds to a timescale $\sim 10^7$ yr. This effect
cannot possibly be due to dust obscuration, as proposed in
\citet{gbgc93}, \citet{ko06}, and \citet{elmegreen07}. A similar
shift occurs between the first appearance of CO and the peak of
Pa$\alpha$, which does not suffer from dust obscuration
\citep{spestr01}.} By contrast, large-scale dynamical
instabilities can amass material over lengthscales much greater
than the sizes of diffuse atomic-hydrogen clouds in a time
consistent with observational constraints.

In the \citet{parker66} instability, undulations of initially
horizontal (i.e., parallel to the galactic plane) magnetic field
lines allow matter to slide down along field lines under the
action of the vertical component of the galactic gravitational
field. The raised portions of the field lines are thereby relieved
of the confining weight of the gas and rise even higher, until the
tension of the field lines increases sufficiently for a final
equilibrium state to be established \citep{mouschovias74}.
Mouschovias et al. (1974) have argued that the Parker instability
is triggered in spiral arms behind spiral density shock waves,
leading to the formation of cloud complexes in the valleys of the
curved magnetic field lines.

Several consequences of this mechanism for the formation of
interstellar clouds are in excellent, in fact unique, agreement
with observations:
\begin{enumerate}
\item The implosion of individual clouds by shock waves within
these complexes leads to the formation of OB associations and
giant H\textsc{ii} regions aligned along spiral arms `like beads
on a string' and separated by regular intervals of $500 - 1000$ pc,
as observed both in our galaxy and in external galaxies
\citep{westerhout63,kerr63,hodge69,morgan70}.

\item The observed moderate enhancement in synchrotron radiation
from the interarm to the arm region of spiral galaxies
\citep{mkb72,kruit73a,kruit73b,kruit73c} found a natural
explanation in the context of the Parker instability (Mouschovias
et al. 1974; Mouschovias 1975a). The relief of magnetic stresses
by buckling of the field lines in the vertical direction in spiral
arms leads to an increase in the magnetic field strength that is
much smaller than that expected from one-dimensional compression.
Moreover, the tendency for the cosmic-ray pressure to distribute
itself uniformly along magnetic flux tubes means that, after the
instability develops, most of the relativistic particles are found
in the regions where the field is weak, i.e., in the inflated
portions of the flux tubes. In other words, the cosmic-ray and
magnetic-energy densities are not in equipartition on scales
$\lesssim 1$ kpc; there is an anticorrelation between them.

\item The anticorrelation between the cosmic-ray density on the
one hand and the magnetic-field strength and gas density on the
other hand also implies that the intensity of synchrotron
radiation {\it along} spiral arms should not vary significantly,
as observed. This is in sharp contrast with expectations based on
correlation between the field strength and the cosmic-ray density
(e.g., equipartition between magnetic and cosmic-ray energy
densities).
\end{enumerate}

Numerical simulations of the Parker instability with applications
to the interstellar medium have been carried out by \citet{bmp97},
\citet{krj98}, \citet{krj01}, and \citet{kos02}. Several questions
concerning the nonlinear evolution of the instability were
addressed. For example, which mode dominates the nonlinear growth
of the instability on timescales of interest in the interstellar
medium (ISM); how fast the instability grows in the nonlinear
regime; and, how great a density enhancement is generated in the
galactic plane by downflow along the magnetic arches. A common
finding in these papers is that the Parker instability {\it by
itself} cannot be the agent responsible for interstellar cloud
formation; the maximum density enhancement during the nonlinear
evolution is only a factor of $\simeq 2$. Instead, the Parker
instability may be thought of as setting the stage upon which
smaller-scale processes, such as cloud formation, star formation,
and subsequent supernova explosions, act out their individual
roles and collectively contribute to the structure and evolution
of the ISM. However, Mouschovias (1974) and BMP97 also noted that
the Parker instability may be {\em directly} responsible for the
formation of GMCs, or cloud complexes, if it can generate a
sufficient density enhancement near the galactic plane for a
critical pressure \citep{fgh69,kp70} to be exceeded, thereby
initiating a phase transition to a cooler, denser cloud phase.
This point has been re-emphasized by \citet{pj00}.

We follow numerically the nonlinear evolution of the Parker
instability in two spatial dimensions accounting for phase
transitions. In Section \ref{S_formulation}, we present the basic
equations and physics of the instability. In Section
\ref{S_method}, the dimensionless equations are given and the
numerical method of solution is described. The results of the
nonlinear development of the instability with phase transitions
from a warm neutral medium (WNM) to a cold neutral medium (CNM)
are presented in Section \ref{S_results}, followed by a summary
and discussion in Section \ref{S_summary}. Contact between theory
and observations is emphasized.

\section{Formulation of the Problem}\label{S_formulation}

\subsection{Basic Equations}\label{SS_equations}

We consider a conducting fluid of mass density $\rho$, thermal
pressure $P$, and temperature $T$, threaded by a frozen-in
magnetic field $\bb{B}$, in an external gravitational field
$\bb{g}$. The magnetohydrodynamic (MHD) equations describing this
system are
\setcounter{eqold}{\value{equation}}\setcounter{eqbid}{0}\addtocounter{eqold}{1}
\renewcommand{\theequation}{\arabic{eqold}\alph{eqbid}}\addtocounter{eqbid}{1}
\begin{equation}\label{E_continuity}
\D{t}{\rho} + \del\bcdot(\rho\bb{v}) = 0\,,
\end{equation}
\addtocounter{eqbid}{1}
\begin{equation}\label{E_force}
\D{t}{(\rho\bb{v})} + \del\bcdot(\rho\bb{v}\bb{v}) = -\del P +
\rho\bb{g} + \frac{1}{4\pi}(\del\btimes\bb{B})\btimes\bb{B}\,,
\end{equation}
\addtocounter{eqbid}{1}
\begin{equation}\label{E_induction}
\D{t}{\bb{B}} = \del\btimes(\bb{v}\btimes\bb{B})\,,
\end{equation}
\addtocounter{eqbid}{1}
\begin{equation}\label{E_energy}
\D{t}{u} + \del\bcdot(u\bb{v}) = -P\del\bcdot\bb{v} - \rho\mc{L} +
\del\bcdot(\kappa\del T)\,,
\end{equation}
\renewcommand{\theequation}{\arabic{equation}}\setcounter{equation}{\value{eqold}}
where $\bb{v}$ is the velocity and $u$ is the internal energy
density. The thermal pressure is a function of both density and
temperature, $P=P(\rho,T)$, and is determined by a balance of
heating and cooling mechanisms in the ISM. It is obtained from the
internal energy density of the gas through the ideal gas relation,
$P=(\gamma-1)u$, where $\gamma$ is the ratio of specific heats
($=5/3$). In equation (\ref{E_energy}), $\mc{L}=\rho\Lambda -
\Gamma$ is the loss function (i.e., net cooling rate per unit
mass), where $\Lambda = \Lambda(\rho,T)$ and $\Gamma$ are the
cooling and heating rates, respectively; $\kappa$ is the thermal
conductivity (see eq. \ref{E_conductivity} below).

\begin{figure}
\centering
\includegraphics[width=3.2in]{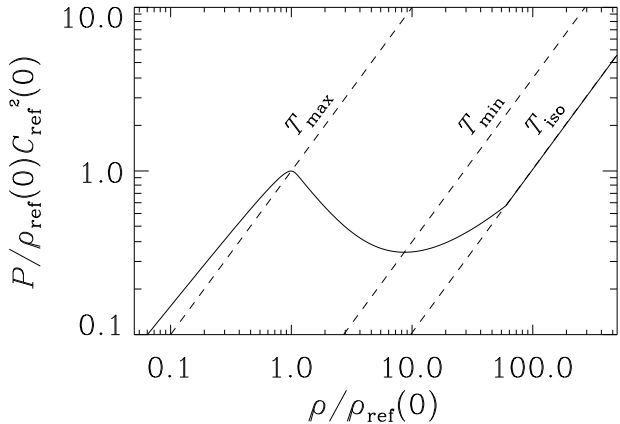}
\caption{Thermal equilibrium curve (solid line) implied by the
heating and cooling functions given by equations \ref{E_heat} and
\ref{E_cool}. The dashed lines are isotherms marking the
transition temperatures between different phases of the ISM; the
WNM corresponds to $T>T_{\rm max} = 5012$ K and the CNM to
$T<T_{\rm min} = 185$ K. The units of $P$ and $\rho$ are their
initial values (in the `reference', stratified state) at the
galactic plane $z = 0$. Typically, $\rho_{\rm ref}(0) = 2.1\times
10^{-24}$ g cm$^{-3}$, corresponding to a number density = 1
cm$^{-3}$, and $C_{\rm ref}(0)$ = 5.73 km s$^{-1}$,
corresponding to a temperature $T_{\rm ref}(0) = 5012$ K (see
\S~3.1).} \label{F_eos}
\end{figure}

We use an equation of state that allows the existence of two
phases, a warm one with $T\sim 10^4$ K and density $n\sim 0.1 - 1$
cm$^{-3}$ and a cold one with $T\sim 10^2$ K and $n \sim 10 - 100$
cm$^{-3}$. Following \citet{nik06}, we adopt the following heating
and cooling rates:\footnote{The functional form of these heating
and cooling rates has its origin in earlier work by
\citet{pb70} and references therein.}
\setcounter{eqold}{\value{equation}}\setcounter{eqbid}{0}\addtocounter{eqold}{1}
\renewcommand{\theequation}{\arabic{eqold}\alph{eqbid}}\addtocounter{eqbid}{1}
\begin{eqnarray}\label{E_heat}
\mu m_{\rm H}\Gamma &= &2\times 10^{-26}\;{\rm erg\;s}^{-1}\,,\\*
\addtocounter{eqbid}{1}\label{E_cool}
\mu m_{\rm H}\frac{\Lambda(T)}{\Gamma} &= &10^7 \exp\left(\frac{-1.184\times 10^5}{T+1000}\right)\nonumber\\*
                                       &+ &1.4\times 10^{-2} \sqrt{T} \exp\left(\frac{-92}{T}\right)\;{\rm cm}^3\,,
\end{eqnarray}
\renewcommand{\theequation}{\arabic{equation}}\setcounter{equation}{\value{eqold}}
where $\mu=1.27$ is the mean mass per particle in units of the
atomic-hydrogen mass, accounting for 10\% He abundance by number.
The minimum and maximum pressures for the
coexistence of the warm/cold phases in a static equilibrium are
$P_{\rm min}/k_{\rm B} \simeq 1600$ K cm$^{-3}$ and $P_{\rm
max}/k_{\rm B} \simeq 5000$ K cm$^{-3}$. The corresponding
transition temperatures $T_{\rm min} = 185$ K and $T_{\rm max} =
5012$ K define cold ($T < T_{\rm min}$), warm ($T>T_{\rm max}$),
and intermediate temperature ($T_{\rm min}<T<T_{\rm max}$) phases.
Since we do not include the additional phase transition to
molecular hydrogen in our loss function, we enforce isothermality
for any gas element whose number density exceeds $n_{\rm iso} =
60$ cm$^{-3}$. The corresponding temperature of this isothermal
branch is $T_{\rm iso}=51.3$ K. The thermal equilibrium curve
implied by these heating and cooling functions is shown in Figure
\ref{F_eos}. The pressure (ordinate) and density (abscissa) are
normalized to their initial midplane values (see
\S~\ref{SS_conditions} below). The dashed lines denote the $T_{\rm
max}$-, $T_{\rm min}$-, and $T_{\rm iso}$-isotherms.

Note that an investigation of the Parker instability accounting
for phase transitions is not the same as an investigation of the
thermal instability occurring alongside the Parker instability
\citep{ames73}. The lengthscales involved in the two instabilities
are so radically different ($\sim 1$ pc for the thermal
instability and $\sim 1$ kpc for the Parker instability) that the
two instabilities proceed in parallel without much effect of one
on the other. The key element in the possibility of the Parker
instability leading directly to the formation of massive
H\textsc{i} clouds and H\textsc{i} cloud complexes is not so much
the detailed nature of the thermal instability but the fact that a
transition from the warm to the cold phase can take place by {\em
some} means (e.g., even if the equation of state on a $\log P$ -
$\log\rho$ diagram had only a horizontal branch joining the two
phases, instead of the branch with a negative slope).

\section{Method of Solution}\label{S_method}

\subsection{Initial and Boundary Conditions}\label{SS_conditions}

We consider a local Cartesian coordinate system $(x,y,z)$
corresponding to cylindrical polar coordinates $(r,\phi,z)$ in a
galactic disk, with the range of the $x$ (radial) coordinate being
$\ll R$ (the distance from the galactic center) so that curvature
terms can be ignored. The magnetic field is initially in the
azimuthal direction ($y$), which we take to be along a spiral arm.
Because the motions involved in the Parker instability are
predominantly azimuthal, rotation can be ignored (see Shu 1974;
also Mouschovias et al. 1974).

Observations show that the interstellar magnetic field is
predominantly parallel to the galactic plane, although arches of
magnetic field lines rising high above the plane are also revealed
\citep{mf70}. \citet{parker66} assumed a zeroth-order initial
(reference) equilibrium state, which, for simplicity, has field
lines exactly parallel to the galactic plane ($z=0$), so that
$\bb{B}_{\rm ref} = B_{\rm ref}(z)\ey$. The vertical galactic
gravitational field (due to stars) is taken to be constant and to
reverse its direction across the galactic plane; i.e., $\bb{g} =
-g(z)\ez $, where $g(z) = -g(-z) = g = $ constant $> 0$. More
complicated gravitational fields have been considered analytically
by \citet{kellman72}, \citet{gs93}, \citet{khr97}, \citet{kh98},
and \citet{kfhsm00} and numerically by \citet{skfmhr00} and
\citet{fkah02}. The main effect of a non-constant gravitational
acceleration is to increase the growth rate of the instability and
shorten the separation of the magnetic valleys.

Although we regard the Parker instability as a driven (or forced)
instability in galactic spiral arms and, therefore, the `initial'
state is not an equilibrium one, we nevertheless use Parker's
initial equilibrium state as a reference state, from which we
obtain the input parameters of the problem. This approach ensures
that (i) any `final' outcome of the evolution of the
gas-field-gravity system can be reached from Parker's stratified
equilibrium state through continuous deformation of the field
lines; and, (ii) the results of the present calculation can be
compared and contrasted with those of past calculations on the
nonlinear evolution of the Parker instability in the ISM. We thus
take the ratio of magnetic and thermal pressures,
\begin{equation}
\alpha \equiv B^2_{\rm ref} / 8\pi P_{\rm ref}\,,
\end{equation}
to be constant in the reference equilibrium state, in which the
density is given by
\begin{equation}\label{E_initstate}
\rho_{\rm ref}(z) = \rho_{\rm ref}(0) \exp\left(-\frac{|z|}{H}\right)\,,
\end{equation}
where
\begin{equation}\label{E_scaleheight}
H \equiv (1+\alpha) C^2_{\rm ref}(0)/ g
\end{equation}
is the {\em total} scale height of the gas in the reference state
and $C_{\rm ref}(0)$ is the isothermal sound speed in the
reference state at the galactic plane. With $T_{\rm ref}(0)= 5012$
K, $g= 3\times 10^{-9}$ cm s$^{-2}$, and $\alpha\simeq 1$, a value
consistent with observations, one finds that $H\simeq 71$ pc. The
magnetic field strength in this reference state is given by
\begin{equation}
B_{\rm ref}(z) = B_{\rm ref}(0)\exp\left(-\frac{|z|}{2H}\right)\,.
\end{equation}
The corresponding pressure and temperature distributions are
determined by requiring the internal energy to satisfy local
thermal equilibrium, based on the heating and cooling functions
given in Section \ref{SS_equations}. For simplicity, we ignore the
effect of cosmic rays in this study, although their effects on the
evolution of the Parker instability are well known by both
analytic arguments \citep{mouschovias75b,mouschovias96,hl00} and
numerical work \citep{hl03,rkhj03,knk04}. Mainly, they tend to
speed up the instability.

The solution of the linear instability problem can either have
field lines with reflection symmetry about $z=0$
(midplane-symmetric or `odd' mode) or allow field lines to cross
$z=0$ (midplane-crossing or `even' mode). In the latter case, the
quantities $\rho$, $\bb{B}$, and $\bb{v}$ are invariant with
respect to a reflection about $z=0$ followed by a translation in
the $y$-direction by an amount $\pm\lambda_{y}/2$, where
$\lambda_{y}$ is the horizontal (along the field lines) wavelength
of the perturbation. (The terms `even' and `odd' refer to the
algebraic form of the velocity perturbation imposed on the
reference equilibrium state.) The `odd' mode in an isothermal
system does not lead to any appreciable density enhancement, and
is therefore less likely than the `even' mode to produce
interstellar clouds in a nonisothermal system. Moreover, BMP97
found that the `even' mode is naturally selected by the nonlinear
evolution of the system. Hence, we study a model with an initial
perturbation that has, and at later times leads to evolution that
preserves, the symmetry of the even mode.\footnote{The simulations
by Kim et al. (1998, 2001) could only study the `odd' mode, due to
their assumed symmetry about the galactic plane. BMP97,
\citet{skfmhr00}, and Kim et al. (2002) considered both the `even'
and `odd' modes.} This symmetry allows us to study a region of
horizontal extent $Y = \lambda_{y}/2$.

The equations are put in dimensionless form by choosing units
natural to the problem. The units of velocity $[v]$, density
$[\rho]$, and magnetic field strength [$B$] are, respectively, the
values of the isothermal sound speed $C_{\rm ref}(0)$, density
$\rho_{\rm ref}(0)$, and magnetic field $B_{\rm ref}(0)$ at the
galactic plane.\footnote{One could choose the unit of $B$ to be
$[\alpha 8\pi\rho_{\rm ref}(0)C^2_{\rm ref}(0)]^{1/2}$, but it is
more convenient to use $B_{\rm ref}(0)$; the two are identical for
$\alpha = 1$.} The unit of length [$L$] is the thermal-pressure
scale height $C^2_{\rm ref}(0)/g$. The implied unit of time $[t]$
is $C_{\rm ref}(0)/g$, the sound-crossing time across a
thermal-pressure scale height. While the evolution of the Parker
instability depends on only one free parameter, $\alpha$, specific
values of $\rho_{\rm ref}(0)$, $g$, and $T_{\rm ref}(0)$ must be
chosen to put the loss function $\mc{L}$ in dimensionless form. We
choose $\rho_{\rm ref}(0) = 2.1\times 10^{-24}$ g cm$^{-3}$
(corresponding to a number density of 1 cm$^{-3}$), $g = 3\times
10^{-9}$ cm s$^{-2}$, and $T_{\rm ref}(0) = 5012$ K. (This
corresponds to a thermally stable state in the warm phase.)
Choosing the free parameter $\alpha = 1$, we find the following
values for the units: [$v$] = 5.73 km s$^{-1}$, [$L$] = 35.45 pc,
[$t$] = 6.05 Myr, and [$B$] = 4.17 $\mu$G.

We choose $Y=12\times[L] = 425.4\,{\rm pc}$, so that
$\lambda_{y}\simeq\lambda_{y,{\rm max}}$ (the horizontal
wavelength corresponding to the maximum growth rate of the linear
instability). The density and velocity are assumed to have
reflection symmetry about the planes $y=0$ and $y=Y$, where the
field lines remain locally horizontal. The upper and lower
boundaries, where the field lines remain undeformed, act as `lids'
to the system, and are placed at $z=\pm Z = \pm 25\times[L]=\pm
886.25\,{\rm pc}$, so that they are far enough from the galactic
plane not to affect the evolution of most of the matter between
them.\footnote{Several numerical studies of the Parker instability
miss this crucial point. For example, the simulations by Kim et
al. (2002) and \citet{ko06} have a total vertical box size of only
$8H$ (with $\alpha = 1/2$). Our total vertical box size, by
contrast, is $\simeq 25H \simeq 1.8$ kpc. A small vertical size of
the computational box suppresses the instability [see eq. (32) of
\citet{mouschovias74}; also, eq. (22) of the review by
\citet{mouschovias96}].} (In the reference state the density at
$z=\pm Z$ decreases to $\exp(-12.5) = 3.73\times 10^{-6}$ of its
value at the galactic plane.) The velocity perturbation is taken
to have the form
\begin{equation}
\delta v_z(y,z) = -\varepsilon\,C_{\rm ref}(0)\cos\left(\frac{\pi y}{Y}\right)
\cos\left(\frac{\pi z}{2Z}\right)\,,
\end{equation}
where $\varepsilon = 0.2$, so that the kinetic energy of the
perturbation is less than 1\% of any other form of energy
(gravitational, thermal, or magnetic) in the system. This choice
of the amplitude of the perturbation does not, of course,
represent the highly nonlinear disturbance introduced in nature by
a spiral density shock wave, and it also has the consequence that
the instability takes a longer time to enter the nonlinear regime
than it does in nature. This choice, nevertheless, allows us to
follow the development of the linear phase of the instability as
well as avoid the introduction of an arbitrariness in the problem
by using such large perturbations that dominate from the outset
the system's evolution. To mimic the natural phenomenon of the
instability driven by a spiral density shock wave without having
to resort to a three-dimensional calculation, we set the origin of
time ($t = 0$) at the instant the linear phase of the evolution
has led to an increase of the {\em maximum} density in the system
by a factor of 2. Times earlier than this instant are assigned
negative values in what follows.

\subsection{The Dimensionless Equations}

We denote the dimensionless time by $\tau$ and the dimensionless
coordinates $(y,z)$ by $(y',z')$, so that
\[
\D{t}{}\rightarrow \frac{g}{C_{\rm ref}(0)}\D{\tau}{}\qquad{\rm and}
\qquad\del\rightarrow \frac{g}{C^2_{\rm ref}(0)}\del'\,.
\]
All other dimensionless quantities are adorned with a tilde. The
dimensionless equations are then
\setcounter{eqold}{\value{equation}}\setcounter{eqbid}{0}\addtocounter{eqold}{1}
\renewcommand{\theequation}{\arabic{eqold}\alph{eqbid}}
\addtocounter{eqbid}{1}
\begin{equation}\label{E_continuity_dimless}
\D{\tau}{\widetilde{\rho}} + \del'\bcdot(\widetilde{\rho}\widetilde{\bb{v}}) = 0\,,
\end{equation}
\addtocounter{eqbid}{1}
\begin{equation}\label{E_force_dimless}
\D{\tau}{(\widetilde{\rho}\widetilde{\bb{v}})} + \del'\bcdot(\widetilde{\rho}
\widetilde{\bb{v}}\widetilde{\bb{v}}) = -\del'\widetilde{P}
- \widetilde{\rho}\,\ez -\alpha\del'\widetilde{B}^2
+ 2\alpha\widetilde{\bb{B}}\bcdot\del'\widetilde{\bb{B}}\,,
\end{equation}
\addtocounter{eqbid}{1}
\begin{equation}\label{E_induction_dimless}
\D{\tau}{\widetilde{\bb{B}}} = \del'\btimes(\widetilde{\bb{v}}\btimes\widetilde{\bb{B}})\,,
\end{equation}
\addtocounter{eqbid}{1}
\begin{equation}\label{E_energy_dimless}
\D{\tau}{\widetilde{u}} + \del'\bcdot(\widetilde{u}\widetilde{\bb{v}}) =
-\widetilde{P}\del'\bcdot\widetilde{\bb{v}} - \widetilde{\rho}\widetilde{\mc{L}} +
\del'\bcdot(\widetilde{\kappa}\del'\widetilde{T})\,,
\end{equation}
\renewcommand{\theequation}{\arabic{equation}}
\setcounter{equation}{\value{eqold}}
where $\widetilde{P} = (\gamma-1)\widetilde{u} = \widetilde{\rho}\,\widetilde{T}$
is the dimensionless pressure, and
\begin{equation}
\widetilde{u} \equiv \frac{u}{\rho_{\rm ref}(0)C^2_{\rm ref}(0)}\,,\;
\widetilde{\mc{L}} \equiv \frac{\mc{L}}{gC_{\rm ref}(0)}\,,\;
\widetilde{\kappa} \equiv \frac{\kappa gT_{\rm ref}(0)}{\rho_{\rm ref}(0)C^5_{\rm ref}(0)}\,,
\end{equation}
are the dimensionless internal energy, loss function, and
conductivity, respectively. The dimensionless density, magnetic
field, and velocity perturbations in the reference state are,
respectively,
\setcounter{eqold}{\value{equation}}\setcounter{eqbid}{0}\addtocounter{eqold}{1}
\renewcommand{\theequation}{\arabic{eqold}\alph{eqbid}}
\addtocounter{eqbid}{1}
\begin{equation}
\widetilde{\rho}_{\rm ref}(z') = \exp\left(-\frac{|z'|}{1+\alpha}\right)\,,
\end{equation}
\addtocounter{eqbid}{1}
\begin{equation}
\widetilde{B}_{\rm ref}(z') = \exp\left(-\frac{|z'|}{2(1+\alpha)}\right)\,,
\end{equation}
\addtocounter{eqbid}{1}
and
\begin{equation}
\widetilde{v}_z(y',z') = -\varepsilon\,\cos\left(\frac{\pi y'}{Y'}\right)
\cos\left(\frac{\pi z'}{2Z'}\right)\,.
\end{equation}
\renewcommand{\theequation}{\arabic{equation}}
\setcounter{equation}{\value{eqold}}
These equations are subject to the boundary conditions
\setcounter{eqold}{\value{equation}}\setcounter{eqbid}{0}\addtocounter{eqold}{1}
\renewcommand{\theequation}{\arabic{eqold}\alph{eqbid}}
\begin{eqnarray}
\addtocounter{eqbid}{1}
\widetilde{v}_z(y',\pm Z') = 0\,,&\widetilde{v}_y(0,z') = 0\,,
&\widetilde{v}_y(Y',z') = 0\,,\\
\addtocounter{eqbid}{1}
\widetilde{B}_z(y',\pm Z') = 0\,,&\widetilde{B}_z(0,z') = 0\,,
&\widetilde{B}_z(Y',z') = 0\,,
\end{eqnarray}
\begin{eqnarray}
\addtocounter{eqbid}{1}
\left.\D{y'}{\widetilde{\rho}(y',z')}\right|_{y'=0,Y'} = 0\,,
&&\left.\D{z'}{\widetilde{\rho}(y',z')}\right|_{z'=\pm Z'}=0\,,\\
\addtocounter{eqbid}{1}
\left.\D{y'}{\widetilde{v}_z(y',z')}\right|_{y'=0,Y'} = 0\,,
&&\left.\D{z'}{\widetilde{v}_y(y',z')}\right|_{z'=\pm Z'}=0\,,\\
\addtocounter{eqbid}{1}
\left.\D{y'}{\widetilde{B}_y(y',z')}\right|_{y'=0,Y'} = 0\,,
&&\left.\D{z'}{\widetilde{B}_y(y',z')}\right|_{z'=\pm Z'}=0\,.
\end{eqnarray}
\renewcommand{\theequation}{\arabic{equation}}\setcounter{equation}{\value{eqold}}

\subsection{Numerical Issues}\label{S_numerics}

\begin{figure*}
\centering
\includegraphics[height=2.8in]{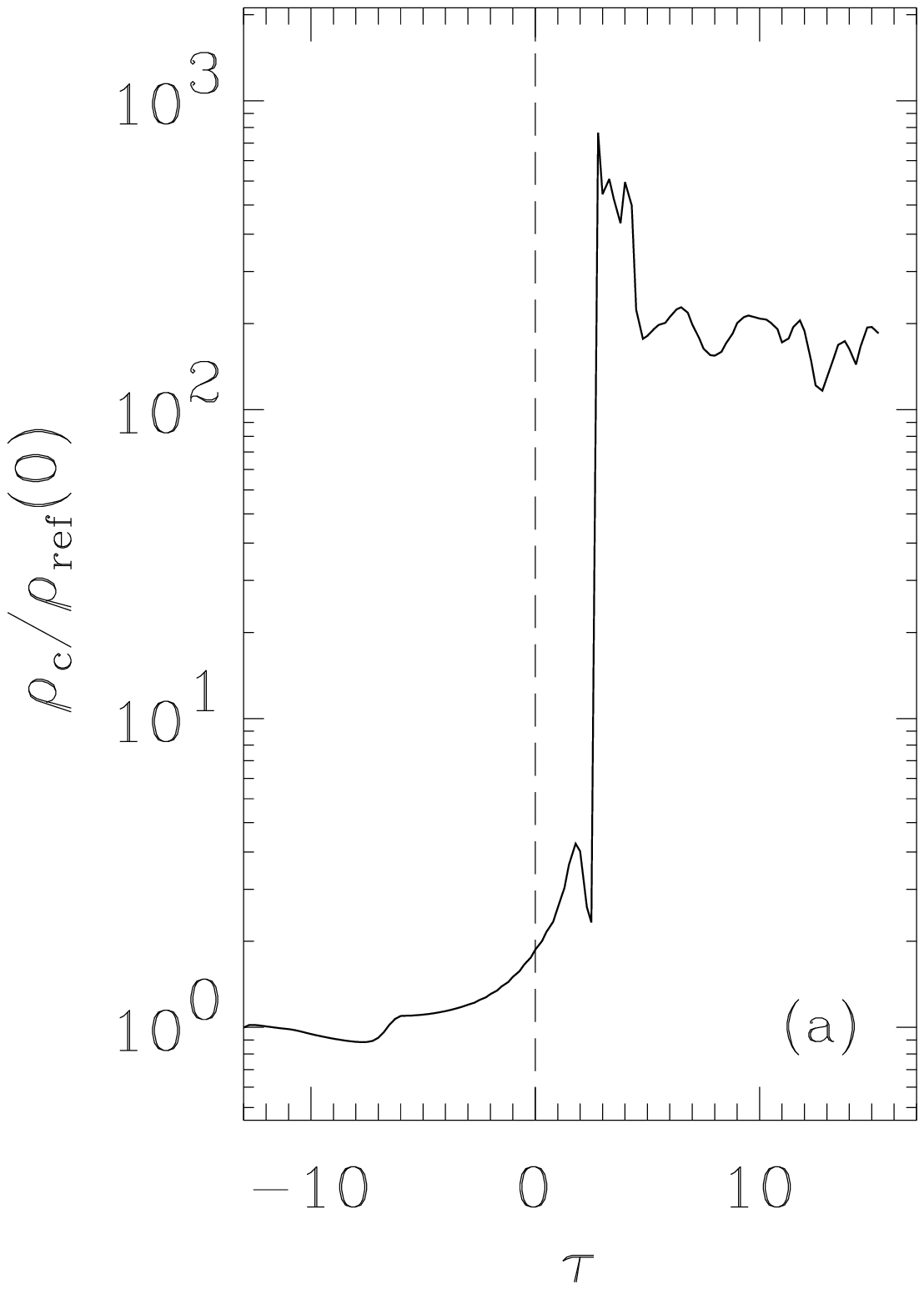}
\qquad
\includegraphics[height=2.8in]{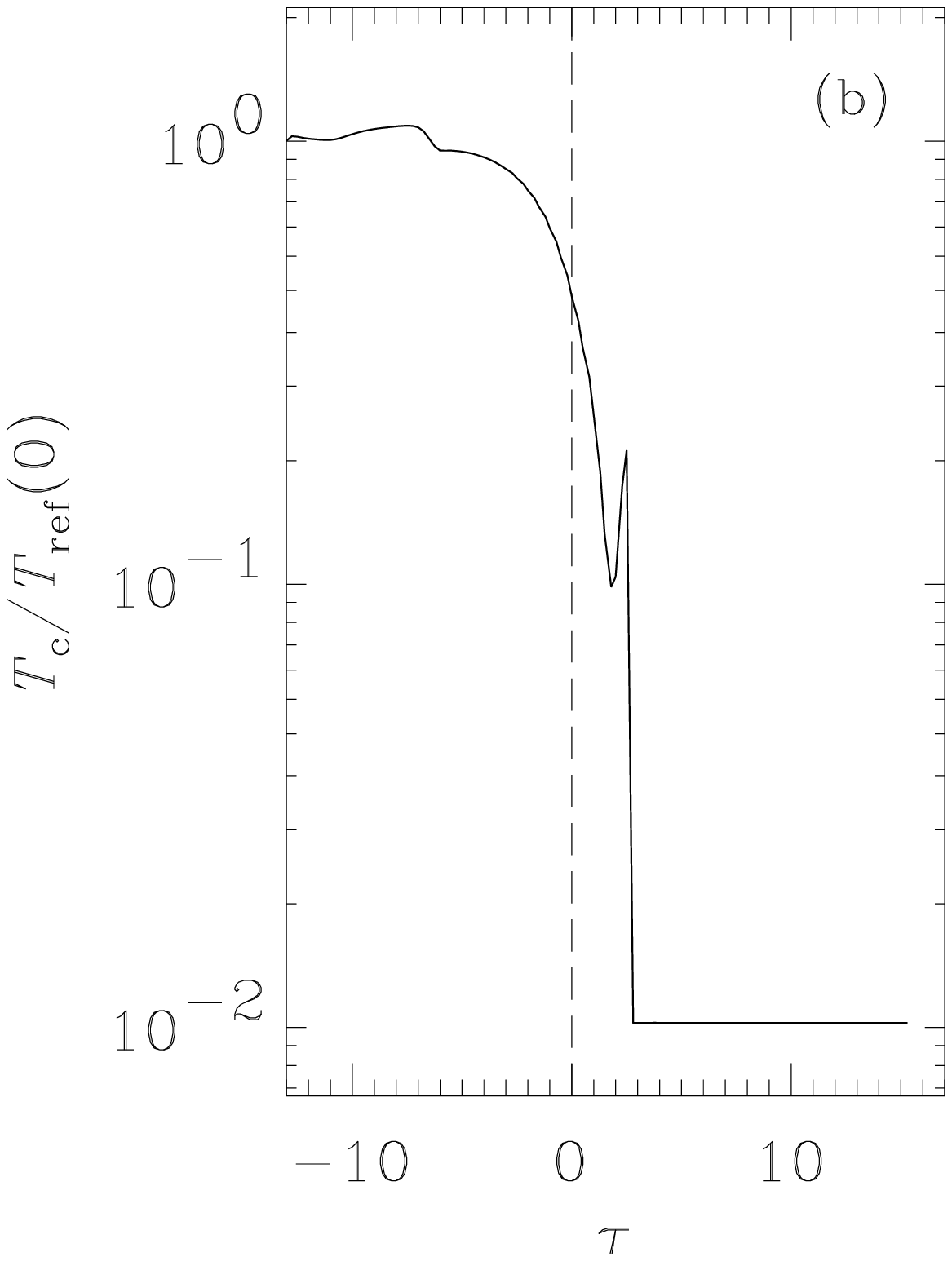}
\qquad
\includegraphics[height=2.8in]{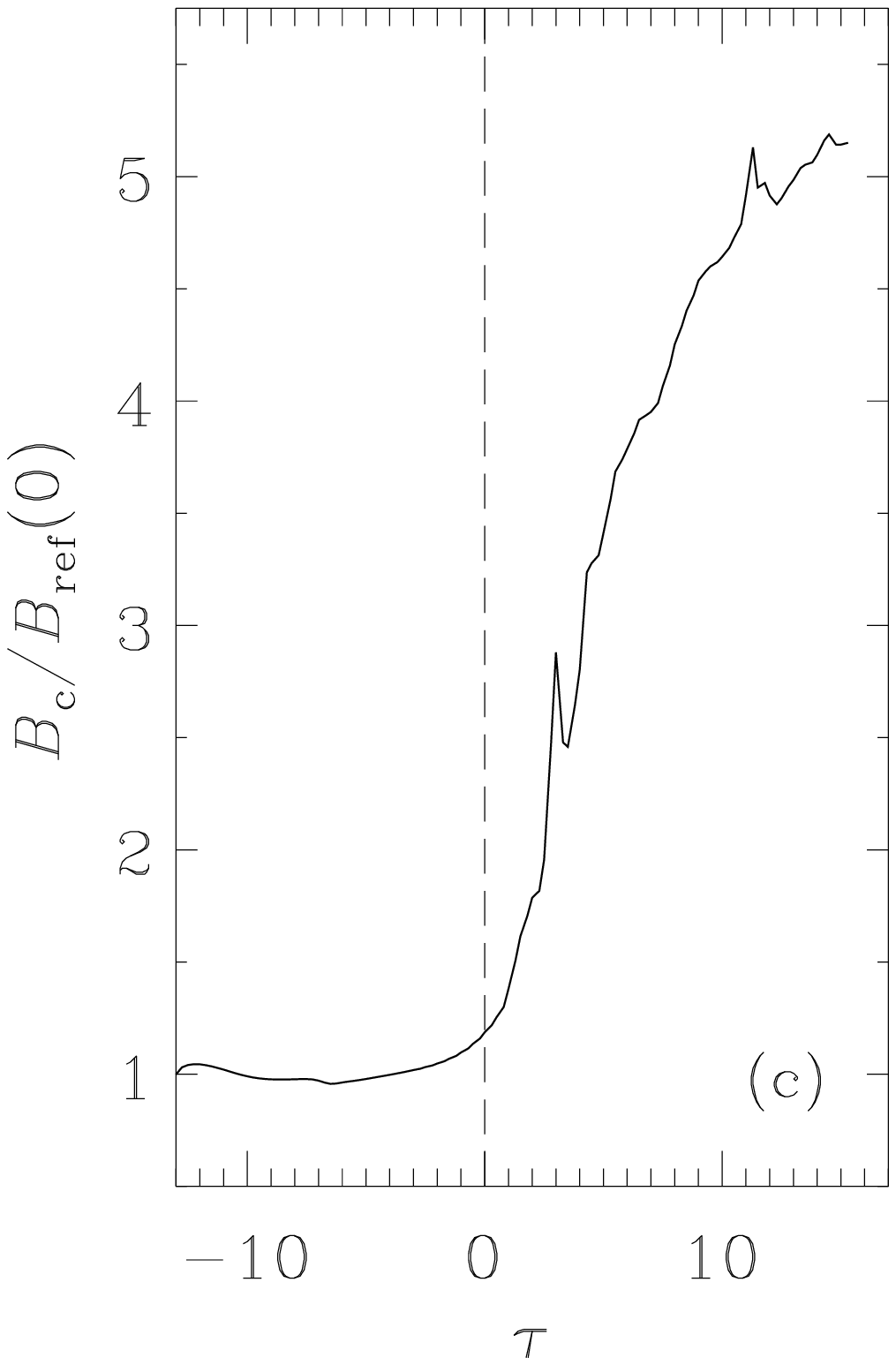}
\newline
\caption{Time evolution of the (dimensionless) central (a) density
$\rho_{\rm c}$, (b) temperature $T_{\rm c}$, and (c) magnetic
field strength $B_{\rm c}$ at $(y',z')=(0,0)$, the bottom of a
magnetic valley. The units of $\rho_{\rm c}$, $T_{\rm c}$, and
$B_{\rm c}$ are their values at the galactic plane, $z' = 0$, in
the initial reference state, namely, $n_{\rm ref}(0) = 1.0$
cm$^{-3}$, $T_{\rm ref}(0) = 5012$ K, and $B_{\rm ref}(0) = 4.17$
$\mu$G. The unit of time is the sound-crossing time across a
thermal-pressure scale height in the reference state, i.e.,
$C_{\rm ref}(0)/g =$ 6.05 Myr.} \label{F_time}
\end{figure*}

We integrate equations (\ref{E_continuity_dimless}) -
(\ref{E_energy_dimless}) using a version of the publicly-available
Zeus-MP code \citep{hnfblcdml06} distributed by the Laboratory for
Computational Astrophysics of the University of California at San
Diego. Zeus-MP uses a time-explicit, operator-split,
finite-difference method for solving the MHD equations on a
staggered mesh, capturing shocks via quadratic and linear
artificial viscosities. Details concerning the algorithms used can
be found in \citet{hnfblcdml06} and references therein.

We have made modifications to this code in order to follow phase
transitions. Because cooling times can be very short compared to
the timescales relevant to cloud formation, the energy equation
update from the net cooling terms is solved implicitly using
Newton-Raphson iteration. At the start of each iteration, the
timestep is initially computed from the Courant-Friedrichs-Levy
condition using the sound speed, Alfv\'{e}n speed, and conduction
parameter (see below). This is followed by a call to the cooling
subroutine. The change in temperature within each zone is limited
to 10\% of its initial value. If this requirement is not met for
all cells in the computational domain, the timestep is reduced by
a factor of 2, and the implicit energy update is recalculated. As
in \citet{po04}, we have found that the timestep is reduced at
most once or twice per convergent timestep.

The update from the conduction operator is solved explicitly,
using a second-order accurate differencing of
$\del\bcdot(\kappa\del T)$. As \citet{ki04} pointed out, the
importance of incorporating conduction in simulations that contain
thermally unstable gas has sometimes been overlooked in past work.
Without conduction, the growth rates for thermal instability are
greatest at the smallest lengthscales, and unresolved growth at
the grid scale may occur. The inclusion of conduction has a
stabilizing effect on the thermal instability at small scales, and
the conduction parameter can be adjusted to allow spatial
resolution of the thermal instability on the computational grid.
We tune the thermal conduction coefficient $\kappa$ so that the
number of zones in a Field length satisfies $\lambda_{\rm
F}/\Delta y = 4$ in the thermally unstable regime:\footnote{The
Field length is $\lambda_{\rm F} = 2\pi\{(\rho^2\Lambda/\kappa
T)[1-(\partial\ln\Lambda/\partial\ln T)]\}^{-1/2}$ when $\Lambda$
is a function of temperature. It is essentially the critical
lengthscale inside which conduction stabilizes the thermal
instability.}
\begin{equation}\label{E_conductivity}
\kappa = \frac{\rho^2\Lambda}{T}\left(\frac{2\Delta y}{\pi}\right)^2
{\rm max}
\left[1-\left(\frac{\partial\ln\Lambda}{\partial\ln T}\right)
,\,0\right]\,.
\end{equation}
As a parcel of gas transits (thermodynamically) from an unstable
state to a stable state, the thermal conductivity vanishes in a
continuous fashion.

The timestep is necessarily limited not only by the usual sound
and Alfv\'{e}n cell-crossing times, but also by a diffusion
timescale that is set by the thermal conduction parameter:
\begin{equation}
(\Delta t)_{\rm cond} = \frac{\rho}{4(\gamma-1)}
\frac{(\Delta y)^2}{\kappa}\,.
\end{equation}
The quadratic dependence on the grid spacing is a general feature
of any numerical problem that includes diffusion. This dependence
is particularly restrictive when high resolution is required to
accurately capture important physical processes. In this work, a
high resolution is needed so that the thermal conduction parameter
is restricted to reasonable values.

The problem is solved on an orthogonal, nonuniform grid. The
horizontal ($y$) grid is spatially uniform, with 2048 zones and a
resolution $\Delta y \simeq 0.21$ pc (i.e., $\Delta y' \simeq
5.9\times 10^{-3}$). The vertical ($z$) grid contains 4096 zones
and is in general nonuniform, with the resolution being greatest
near the galactic plane. There are 2048 uniformly-spaced zones in
the 15 initial thermal-pressure scale heights straddling the
galactic plane, so that $\Delta z \simeq 0.26$ pc (i.e., $\Delta
z'\simeq 7.3\times 10^{-3}$); the remaining 2048 zones are
logarithmically-spaced above and below this uniform-grid region.
The transition between the uniform and nonuniform regions is
chosen so that the grid spacing varies smoothly.

\section{Results}\label{S_results}
\begin{figure*}
\centering
\includegraphics[height=4in]{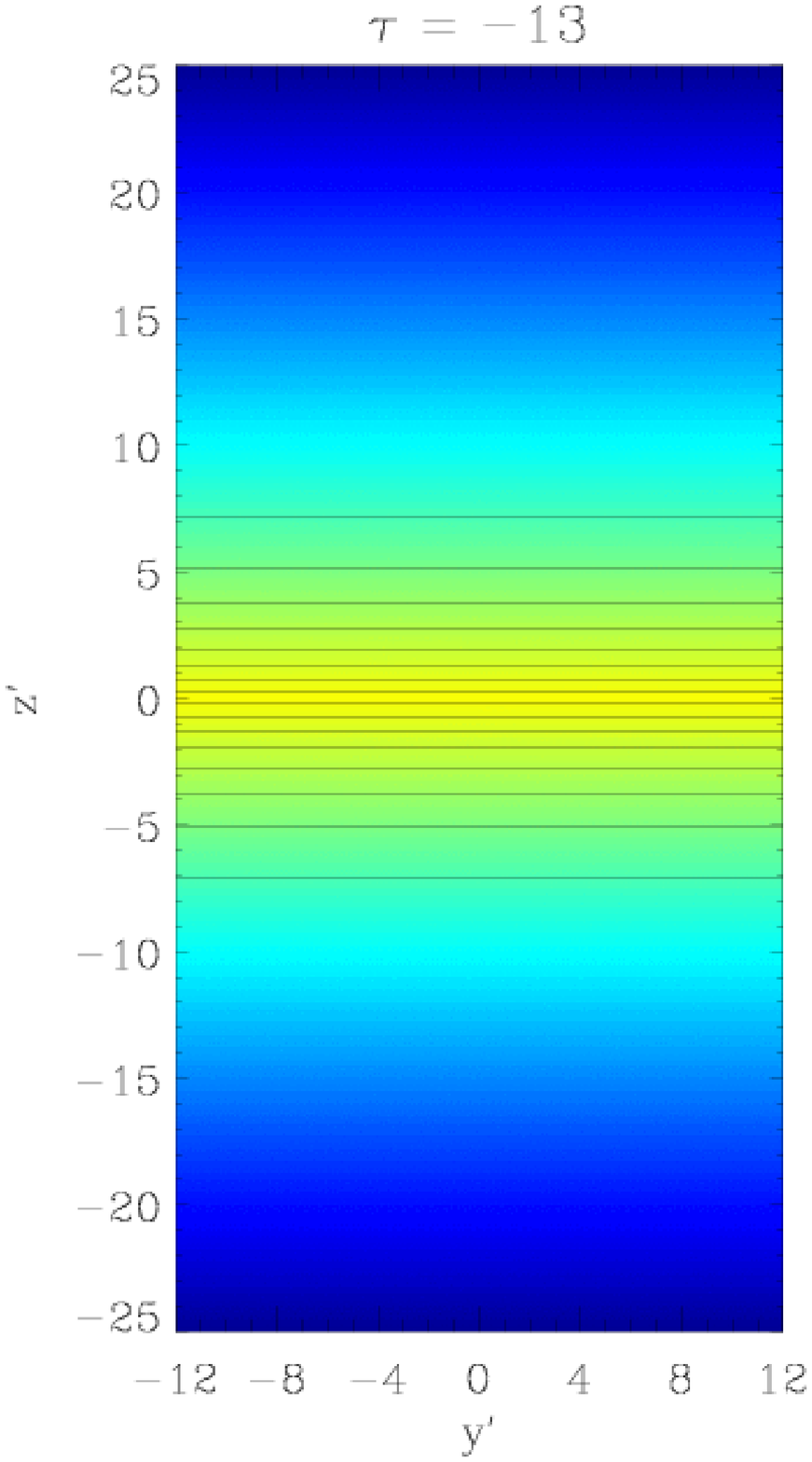}
\quad
\includegraphics[height=4in]{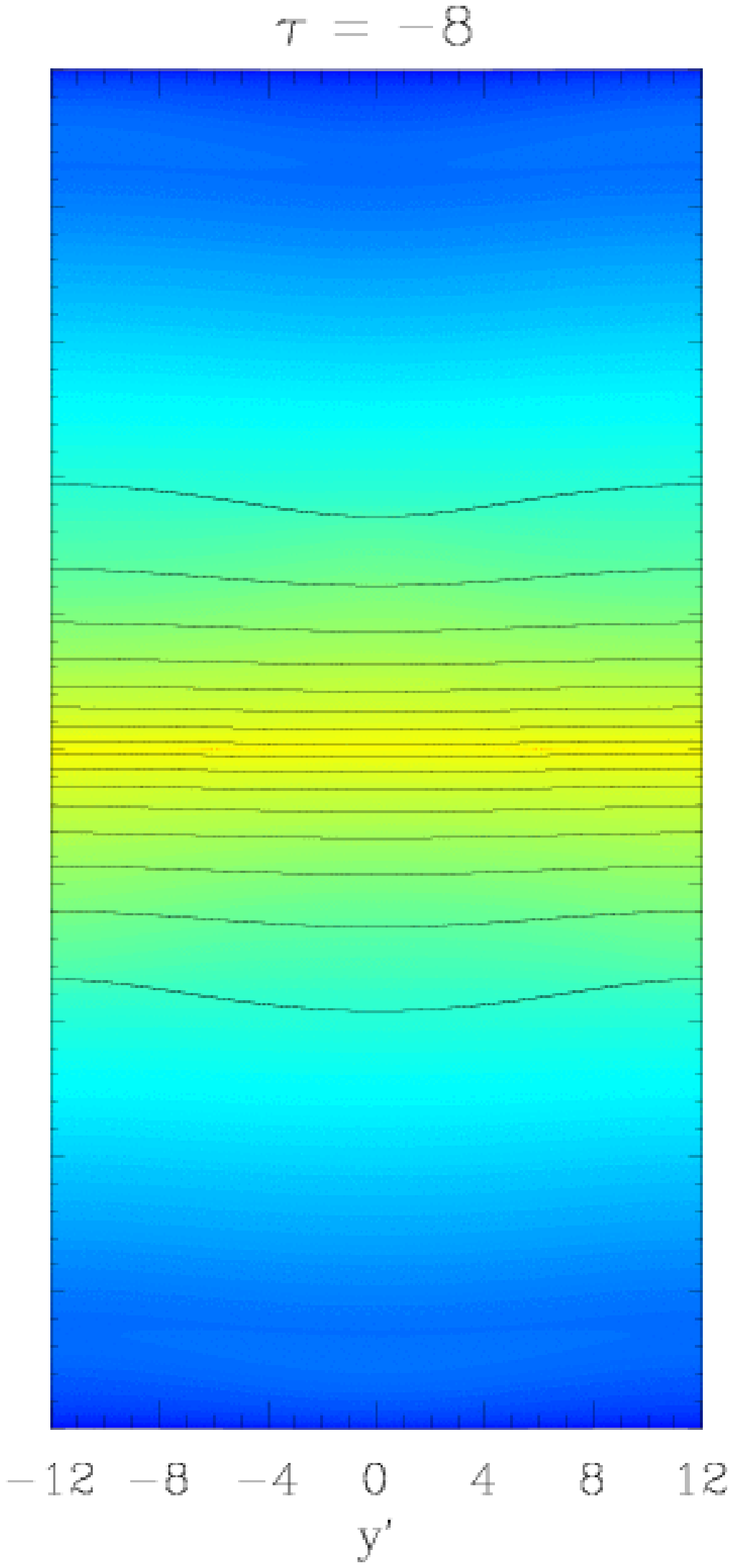}
\quad
\includegraphics[height=4in]{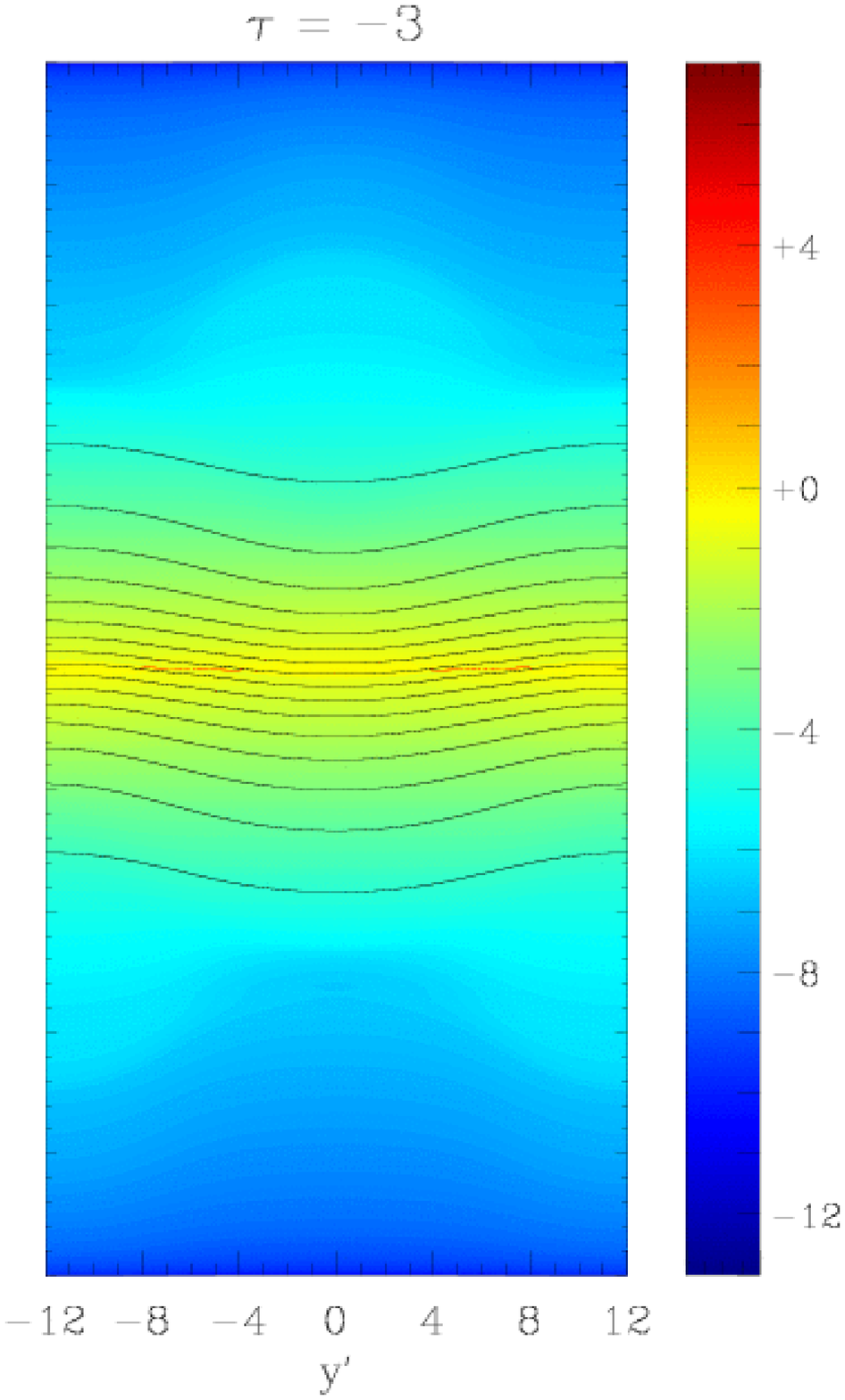}
\newline
\newline
\newline
\includegraphics[height=4in]{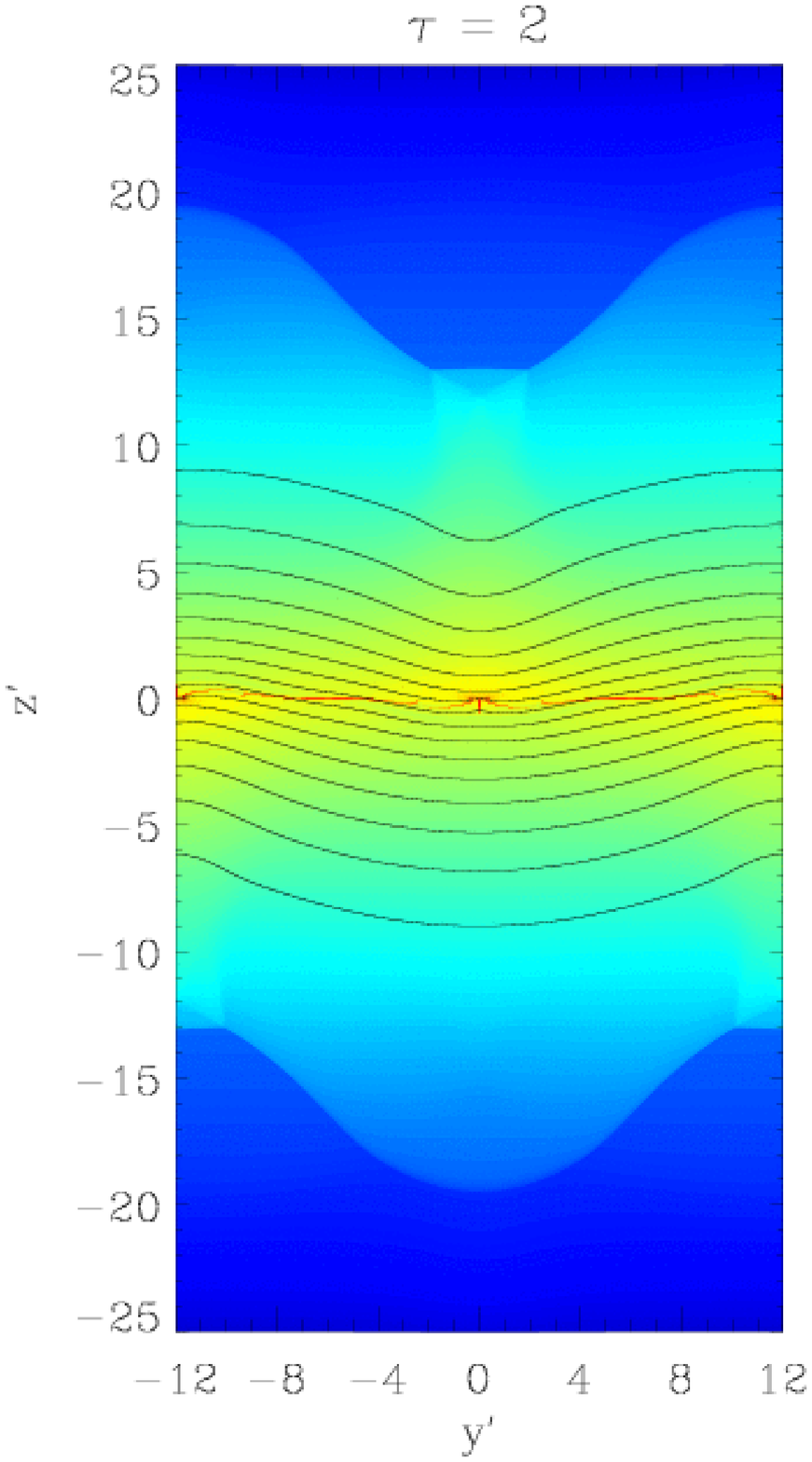}
\quad
\includegraphics[height=4in]{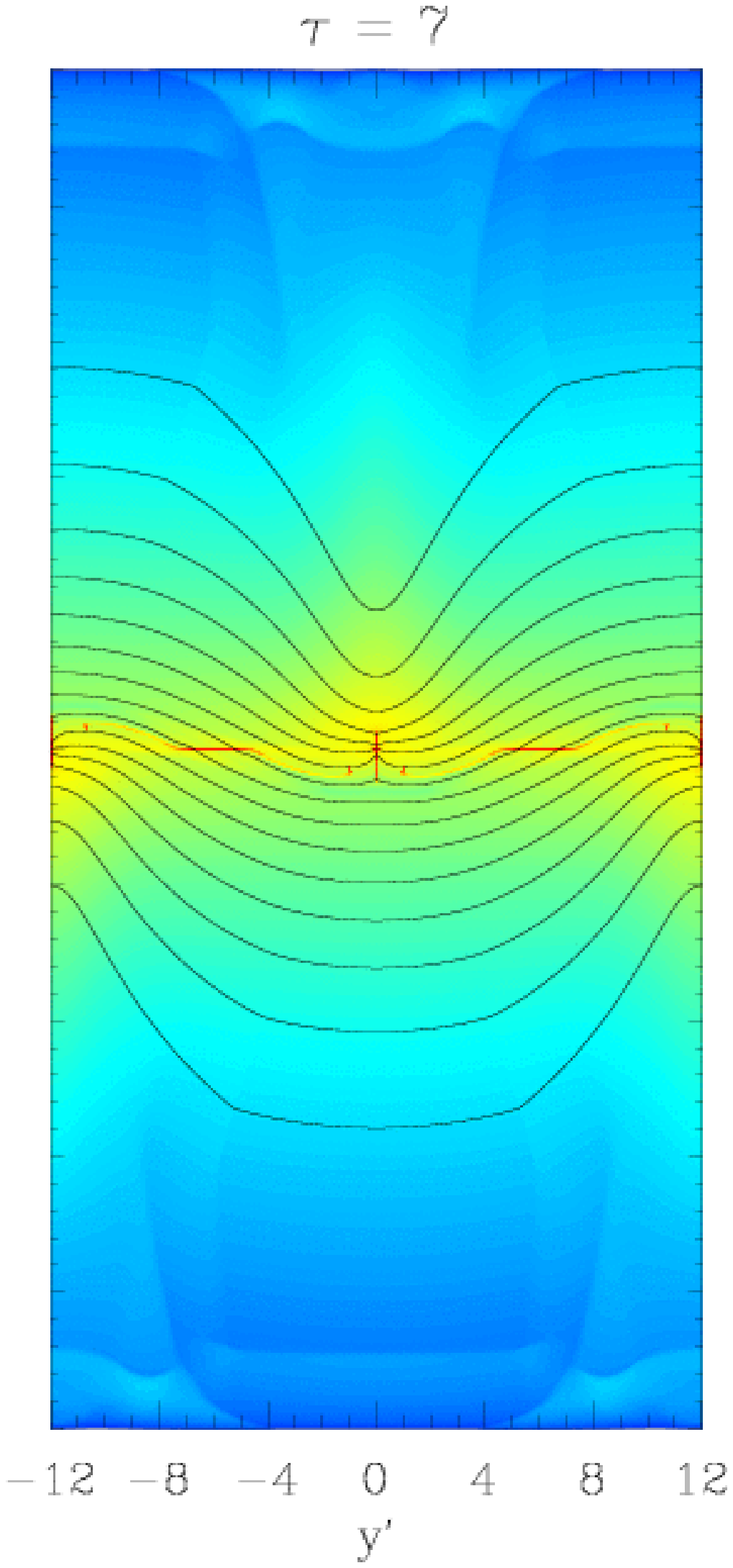}
\quad
\includegraphics[height=4in]{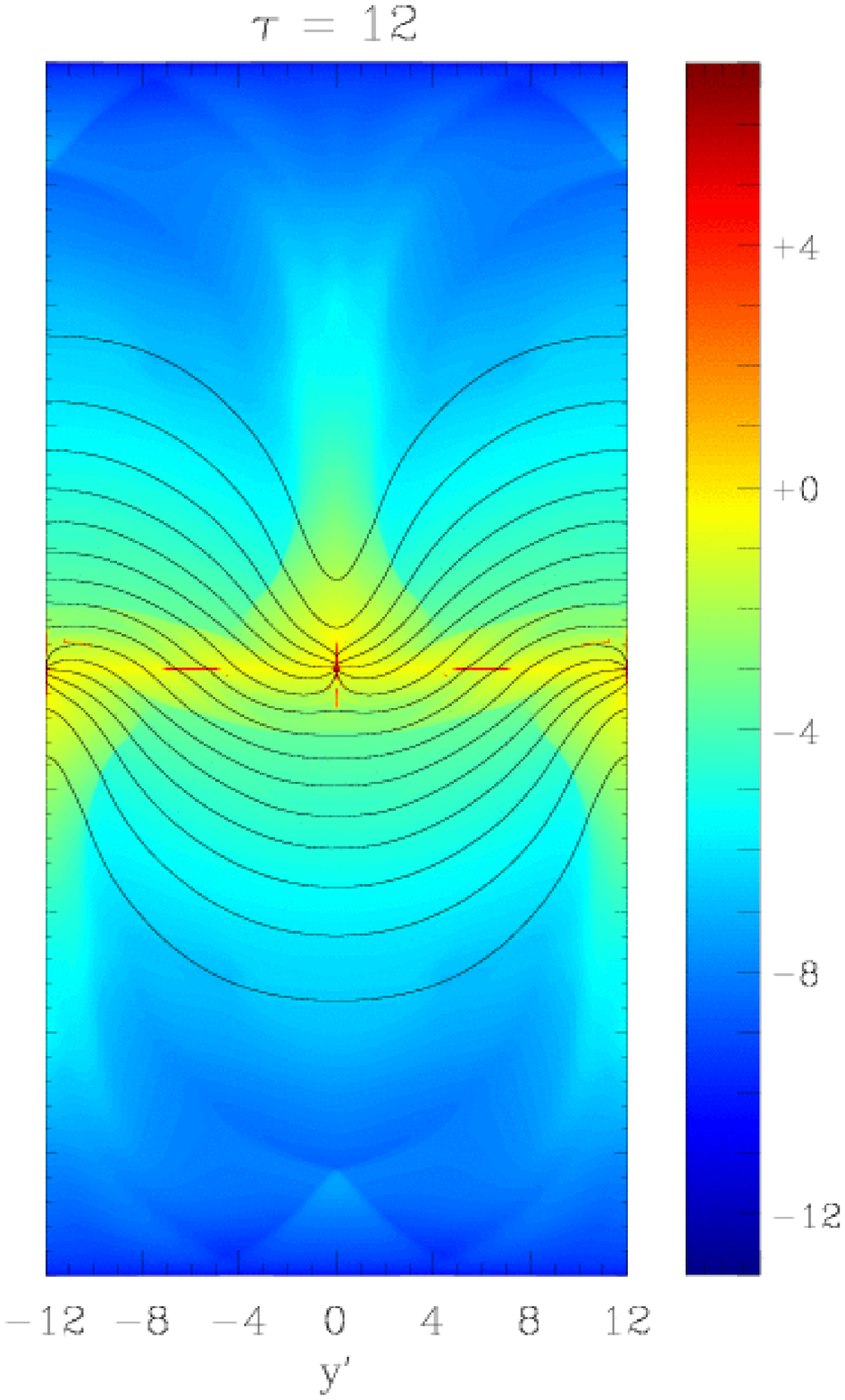}
\newline
\caption{Contour plots of the density (color) and magnetic field
lines (solid lines) at times $\tau = -13$, $-8$, $-3$, 2, 7, and
12 in the ($y', z'$) plane. The colorbar gives the correlation
between the colors and the natural logarithm of the density [e.g.,
`+4' denotes a dimensionless density of $\exp(+4)\simeq 54.6$].
The unit of length is the thermal-pressure scale height in the
reference state, $C^2_{\rm ref}(0)/g =$ 35.45 pc, the unit of
density is $\rho_{\rm ref}(0)= 2.1\times 10^{-24}$ g cm$^{-3}$,
and the unit of time is $C_{\rm ref}(0)/g =$ 6.05 Myr.}
\label{F_contour}
\end{figure*}

In Figure \ref{F_time}, we show the time evolution of the central
(dimensionless) (a) density, (b) temperature, and (c) magnetic
field strength at $(y',z')=(0,0)$, which is located at the bottom
of a magnetic valley. As found by previous studies of the {\em
isothermal} Parker instability, there are three distinct phases of
evolution: linear, nonlinear, and relaxation into a `final'
equilibrium state.

During the linear phase ($-13 \le \tau \le 0$), the velocities
remain subsonic and the central density, temperature, and
magnetic-field strength vary only slightly. Soon after the onset
of the nonlinear phase (at $\tau = 0$), the central density,
temperature and magnetic-field strength change very rapidly. After
the density has increased by a factor of $\simeq 2$, the pressure
of many fluid elements has also increased above the critical value
for a phase transition to occur to a cold (cloud) phase. This
leads to a rapid density enhancement by more than two orders of
magnitude in a time $\Delta\tau \lesssim 3$ (i.e., $\Delta
t\lesssim 18$ Myr). During the phase transition, approximately
isobaric evolution causes the temperature to decrease by about two
orders of magnitude. The magnetic field strength increases only
slightly during the early part of the nonlinear phase, since gas
motions are primarily along field lines. However, after enough gas
has accumulated in the valleys of the field lines, its weight
compresses the field lines and the maximum magnetic-field strength
increases by a factor of three (from $\simeq 4\,\mu$G to $\simeq
12\,\mu$G). During the subsequent relaxation phase, the compressed
gas rebounds because it had overshot its equilibrium state, and
eventually settles in a state that has a central density $\simeq
200$ cm$^{-3}$. The magnetic field strength continues to increase
as more mass elements enter the cloud phase and the galactic
gravitational field (i) compresses the cloud perpendicular to the
galactic plane and (ii) flattens it in the galactic plane. The
resulting physical conditions are such that, if we had included
the gas' self-gravity in the calculation, individual
self-gravitating clouds would separate out. The density
oscillations evident after $\tau\simeq 4$ occur about a `final'
equilibrium state; they are the result of motions left over by the
cloud formation process.

As explained in Section 3.1, the linear phase of evolution is
expected to be very short-lived in nature because the instability
is externally driven by a spiral density shock wave (see
Mouschovias et al. 1974). For this reason, the `origin' of time is
chosen to be the instant by which the central density has
increased by a factor of 2. It is thus seen that the time required
to form cold interstellar cloud complexes from the WNM does not
exceed $\simeq 30$ Myr.

Figure \ref{F_contour} exhibits contour plots of the density
(color) and magnetic field lines (solid lines) at times $\tau=
-13$, $-8$, $-3$, 2, 7, and 12. For ease in visualization, we show
a region that covers one full horizontal wavelength, $-Y'\le y'
\le +Y'$, of the instability. Regions of high (low) density are
colored red (blue) with a color continuum between them (as shown
in the colorbar to the right of the figures). By the end of the
linear phase of the evolution ($\tau=0$), the magnetic field lines
have begun to deform significantly and the gas has begun its rapid
descent into the magnetic valleys. The maximum velocity has
exceeded the sound speed by a factor of 1.61. The evolution up to
this point is nearly indistinguishable from the evolution of the
isothermal Parker instability (determined by BMP97), because few
gas elements have transitioned into the stable CNM phase by this
time (see below). By $\tau=2$, the system is well into the
nonlinear stage of its evolution. The field lines have become
considerably arched, even near $z'=0$. There is a strong flow of
gas [maximum velocity $2.63 C_{\rm ref}(0)$] into the magnetic
valleys, leading to the formation of shocks. By $\tau=7$, the
rising portions of the field lines have almost stopped their
expansion and are nearly at their `final' equilibrium positions.

\begin{figure}
\centering
\includegraphics[width=3.2in]{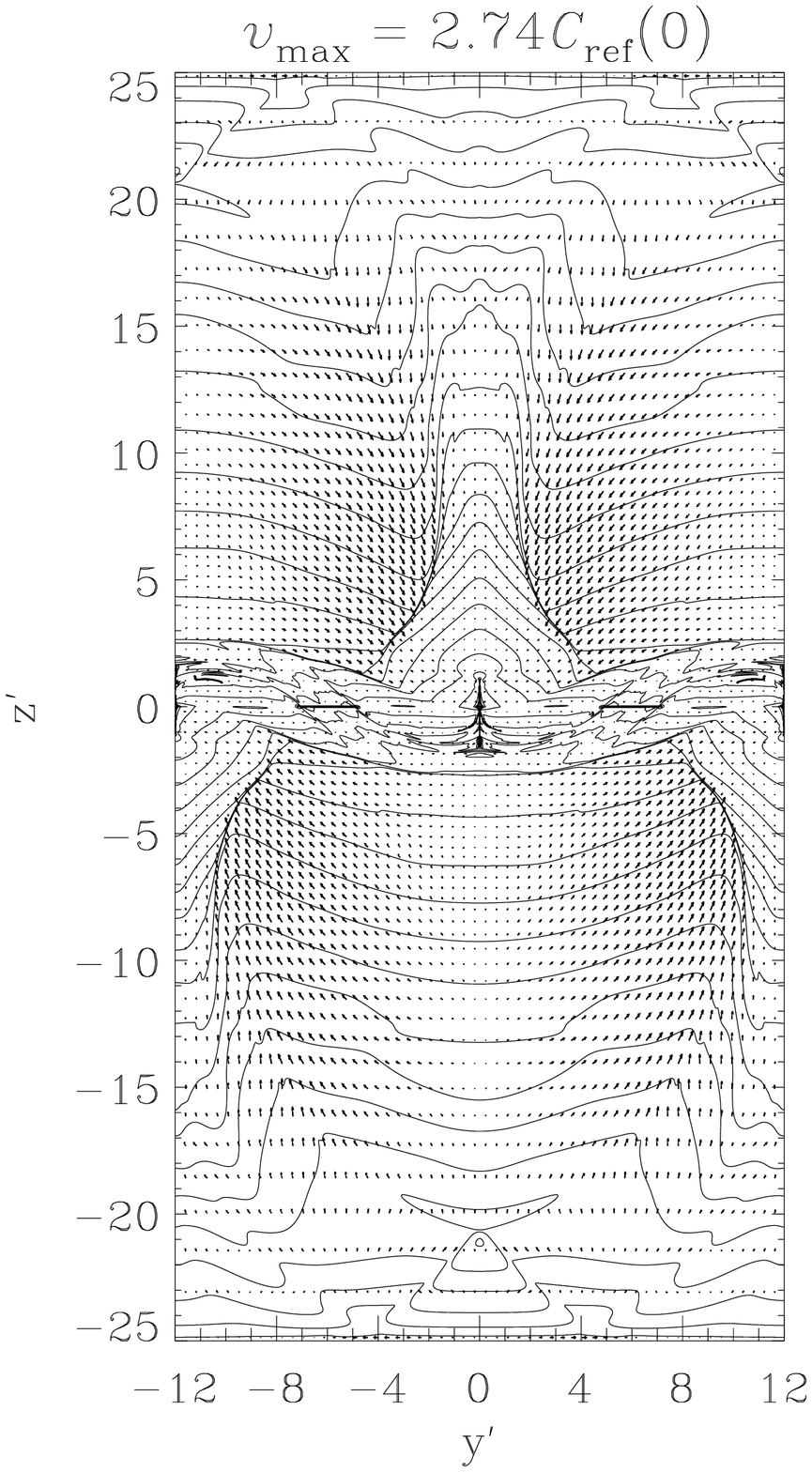}
\newline
\caption{Contour plot of the density (solid lines) at time
$\tau=12$ in the ($y', z'$) plane. The velocity vectors (arrows)
are also shown and the maximum value of the velocity [$=2.74C_{\rm
ref}(0)\simeq 15.7$ km s$^{-1}$] is printed above the figure. The unit of length is
$C^2_{\rm ref}(0)/g =$ 35.45 pc. Note the presence of distinctive
S-shaped shocks extending far above and below the galactic plane.}
\label{F_velocity}
\end{figure}

In Figure \ref{F_velocity}, we display the large-scale velocity
(arrows) and density (solid lines) fields at time $\tau=12$. There
is considerable density structure in the galactic plane. The
velocity field shows prominent and distinctive S-shaped shocks
extending far above and below the galactic plane. Shocks of this
character have also been seen in the BMP97 simulations of the {\em
isothermal} Parker instability. \citet{mouschovias96} noted that,
if such shocks are observed, they cannot possibly be mistaken for
supernova remnants and can therefore be taken as evidence for the
nonlinear development of the Parker instability in the
interstellar medium.

\begin{figure*}
\centering
\includegraphics[height=3.25in]{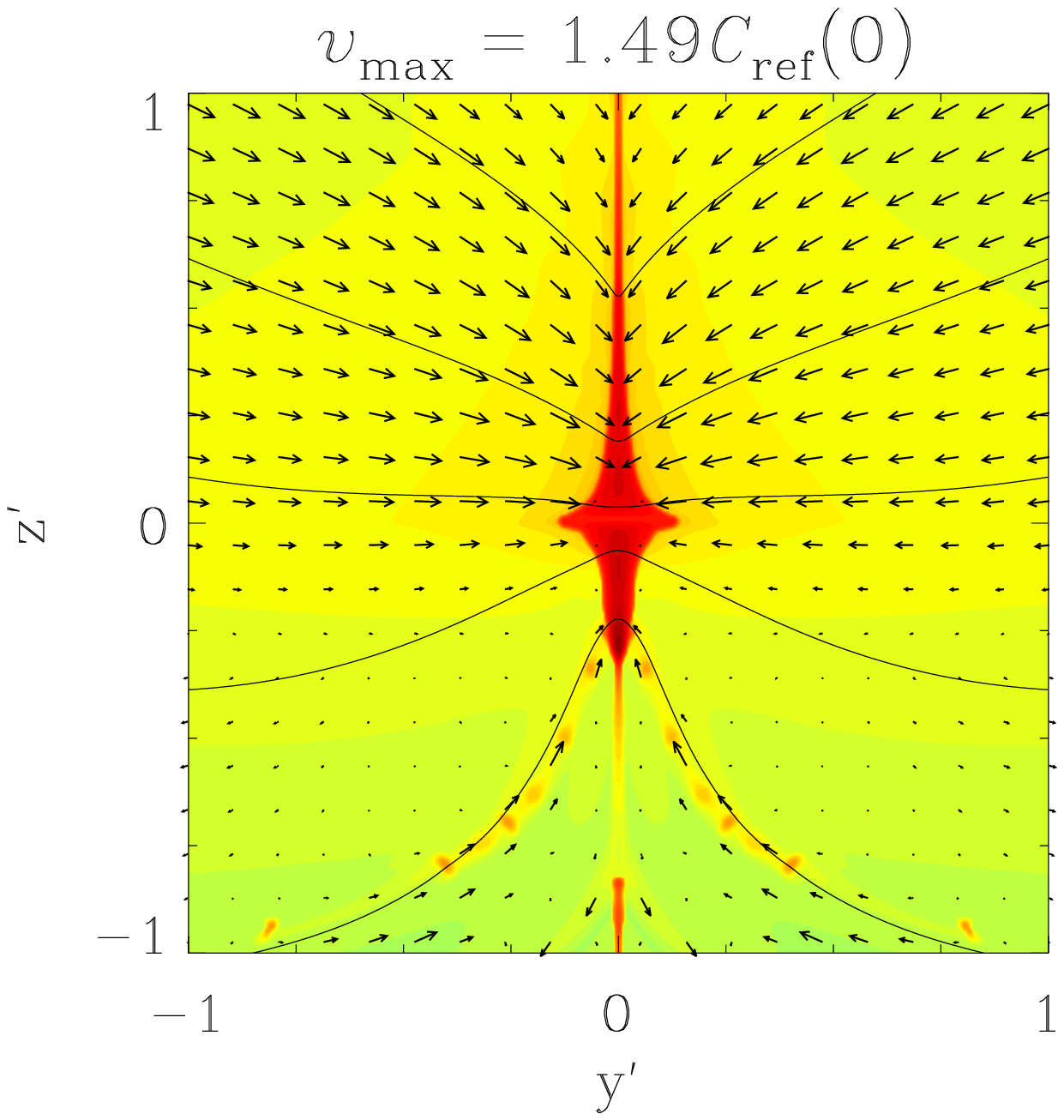}
\qquad\quad
\includegraphics[height=3.25in]{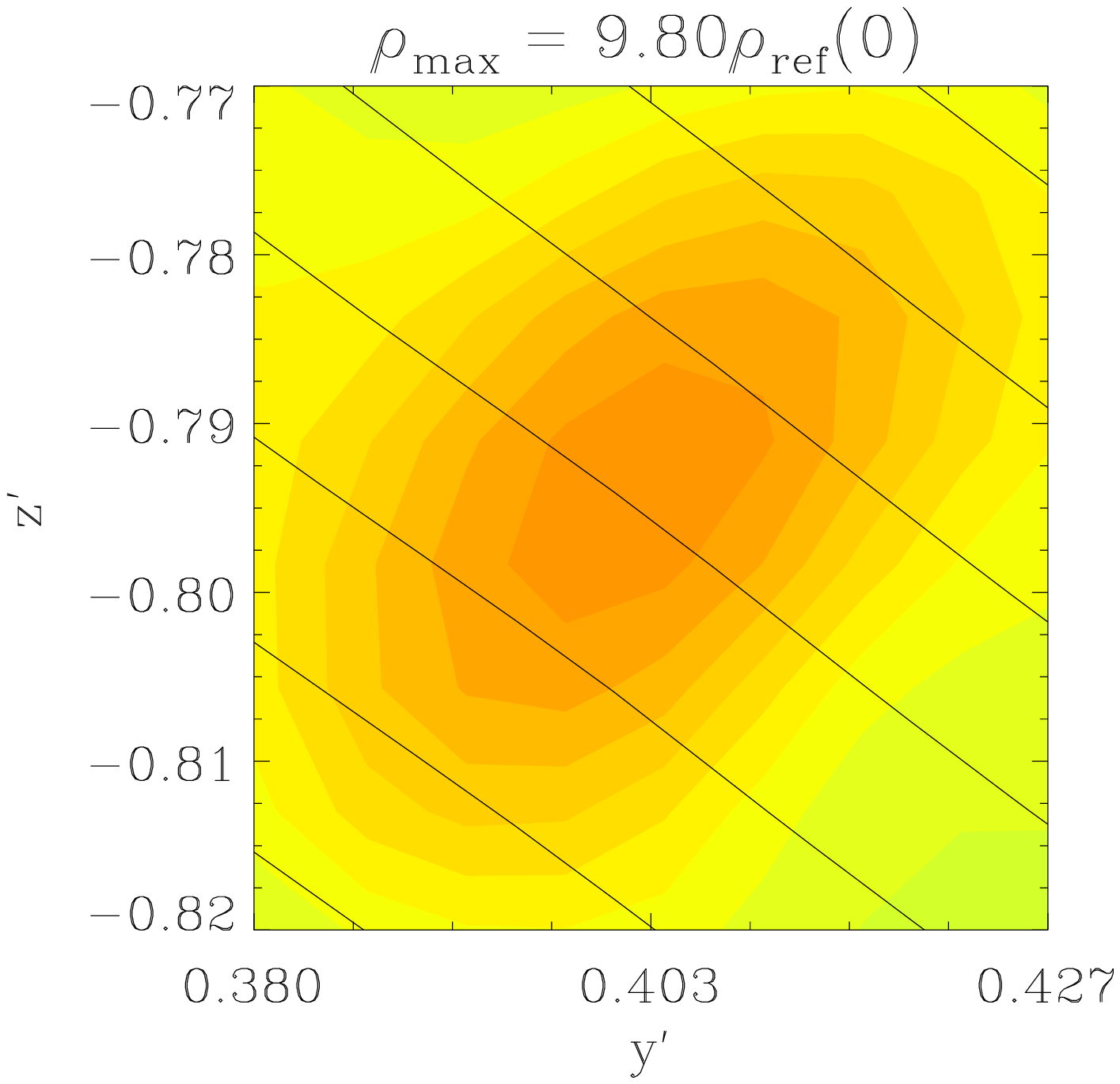}
\newline
\caption{Contour plots of the density (color) and magnetic field
lines (solid lines) at time $\tau=12$ in the ($y', z'$) plane,
zoomed in on (a, {\it left}) a CNM `sheet' in a magnetic valley and (b, {\it right}) a CNM
cloudlet. In (a) the velocity vectors (arrows) are also shown and
the maximum velocity [$=1.49C_{\rm ref}(0) \simeq 8.5$ km
s$^{-1}$] is given. In (b) the maximum density [$=9.80\rho_{\rm
ref}(0) \simeq 2\times 10^{-23}$ g cm$^{-3} \simeq 10 \, {\rm
cm}^{-3}$] of the cloudlet is given. The unit of length is
$C^2_{\rm ref}(0)/g =$ 35.45 pc.}
\label{F_zoom}
\end{figure*}

Figure \ref{F_zoom}a shows the innermost $-1\le y' \le +1$, $-1
\le z' \le +1$ part of the system shown in Figure \ref{F_contour}
at time $\tau=12$. The CNM has a clear sheet-like morphology
extending high above the galactic plane, but it also has a
significant extent in the galactic plane, in agreement with
observations by \citet{hj76} and \citet{ht03}. The maximum
velocity in this region is $v_{\rm max}=1.49C_{\rm ref}(0)\simeq 8.5$
km s$^{-1}$, also in agreement with observations. Gas elements
slide down the deformed magnetic field lines and undergo a phase
transition, from the WNM to the CNM. A shock front marks the
CNM-WNM interface, across which the density changes by a factor
$\simeq 40$ in $\sim 1$ pc. In addition to the large structures
discussed above, there are small cold `cloudlets' of size $\sim 1$
pc strung along field lines near the bottom of the panel. These
resemble the cloudlets discovered by \citet{heiles67} and further
discussed in \citet{ht03}. Figure \ref{F_zoom}b shows a typical
cloudlet; the local magnetic field lines are also shown. The
maximum number density of this cloudlet is $\simeq 10$ cm$^{-3}$;
its major (minor) axis is $\simeq 2$ (1.2) pc, giving an axis
ratio of $\simeq 5:3$. The median column density is $\simeq
0.3\times 10^{20}$ cm$^{-2}$. These numbers are in good
quantitative agreement with the observations discussed
in \S~8.3 of \citet{ht03}.
\begin{figure*}
\centering
\includegraphics[width=3.4in]{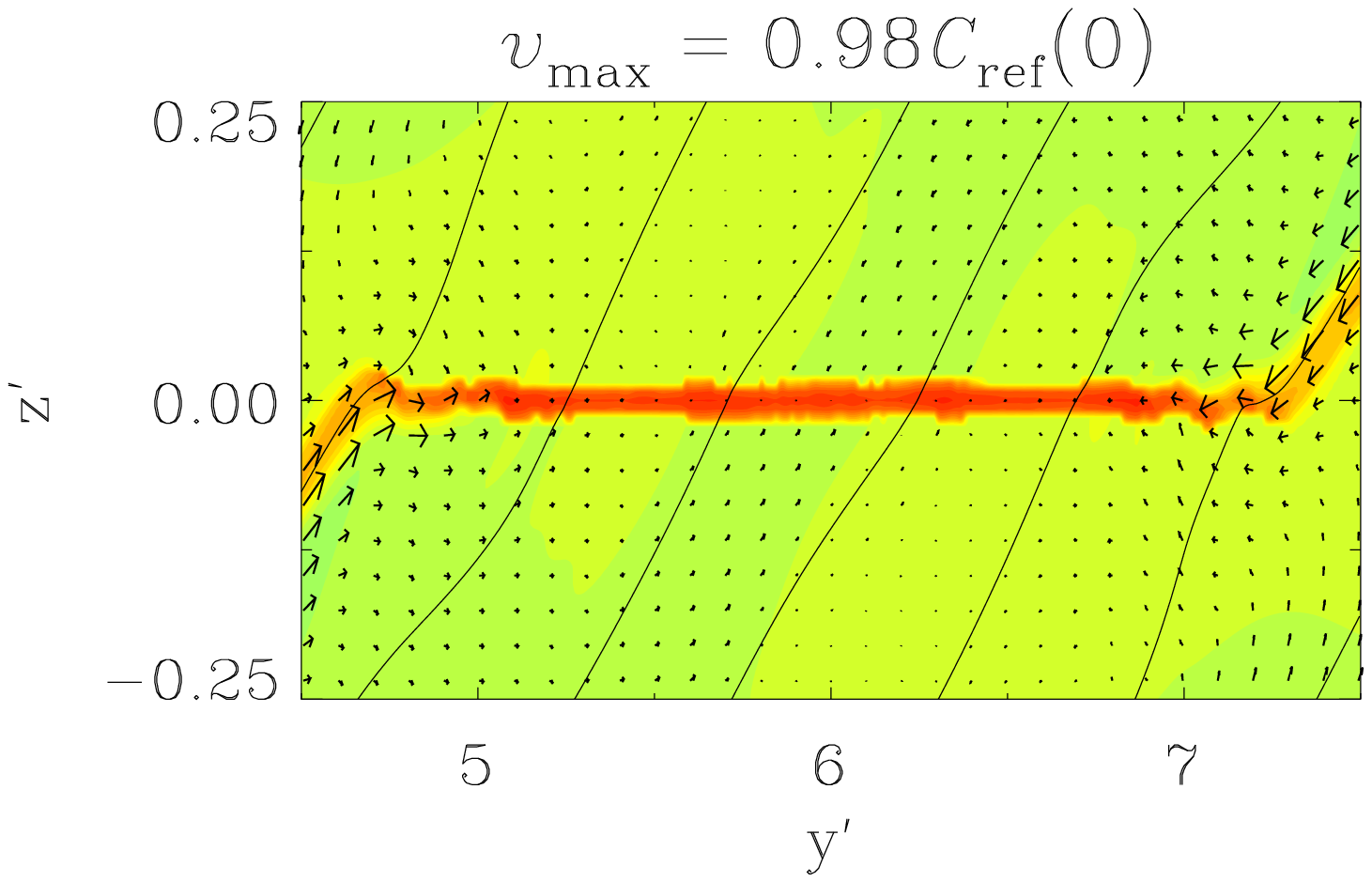}
\quad
\includegraphics[width=3.4in]{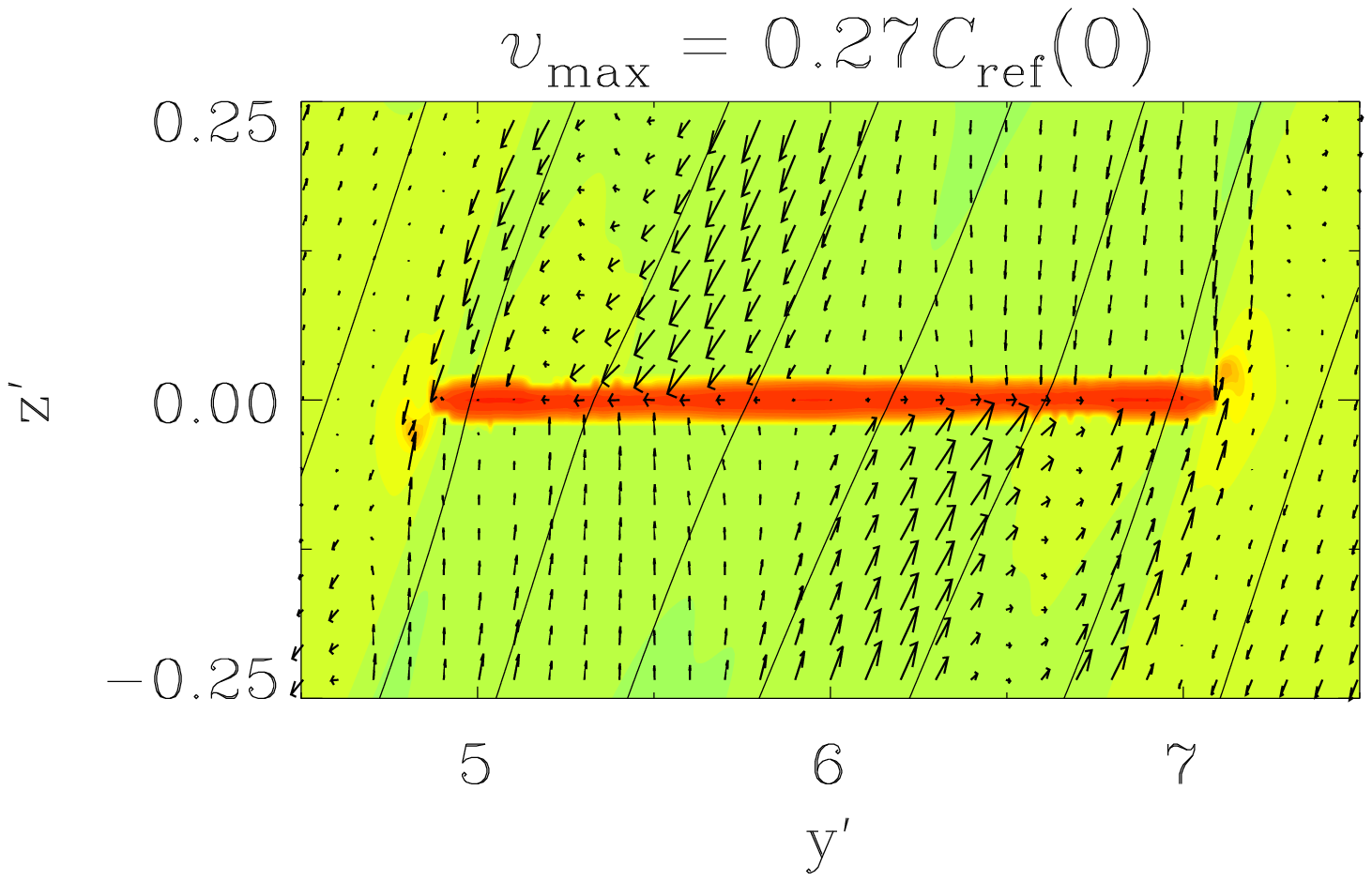}
\newline
\caption{Contour plots of the density (color) and magnetic field
lines (solid lines) in the ($y', z'$) plane, zoomed in on a CNM
`sheet' in the galactic plane, at times (a, {\it left}) $\tau=7$ and (b, {\it right}) $\tau=12$. The velocity vectors (arrows) are also shown and are
normalized to the maximum speed in the region [= (a) $0.98C_{\rm
ref}(0)\simeq 5.6$ km s$^{-1}$; (b) $0.27C_{\rm ref}(0)\simeq
1.5$ km s$^{-1}$]. The unit of length is $C^2_{\rm ref}(0)/g =$
35.45 pc. Note that the $z'$ axis is stretched out of proportion
in order to show the vertical structure of the sheet.}
\label{F_gp}
\end{figure*}

There is also a great deal of CNM away from the magnetic valleys.
In Figure \ref{F_gp}, we show a cold H\textsc{i} cloud elongated
in the galactic plane at times (a) $\tau=7$ and (b) $\tau=12$. There are notable density
variations and ripples along the CNM-WNM interface, where there
is significant velocity shear. The major (minor) axis of the cloud
at $\tau=12$ is $\simeq 80$ (1.35) pc, giving an axis ratio of
$\simeq 60:1$. The magnetic field makes an angle of $\lesssim
22^\circ$ with respect to the minor axis of the cloud.

\begin{figure*}
\centering
\includegraphics[width=3.2in]{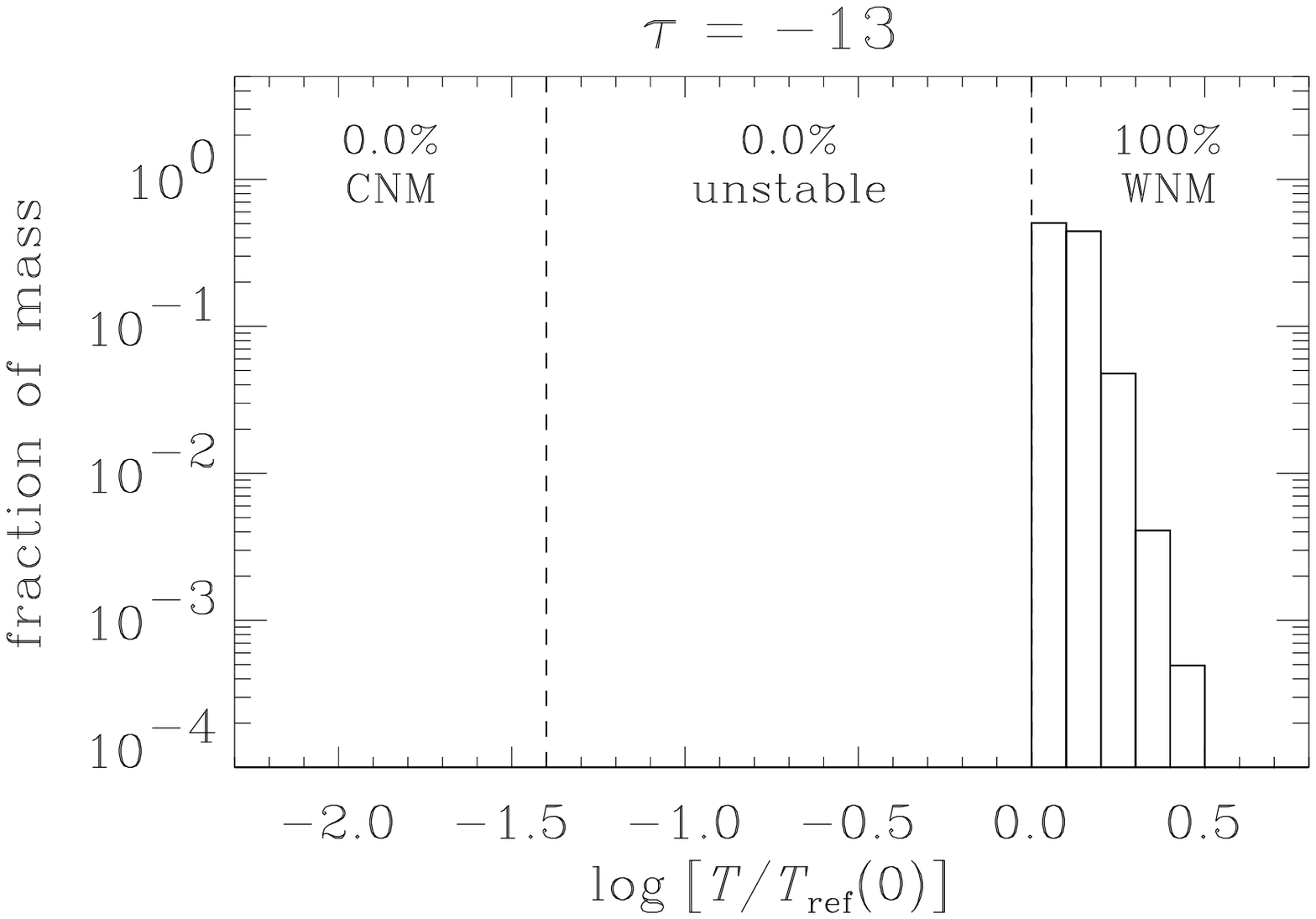}
\qquad\quad
\includegraphics[width=3.2in]{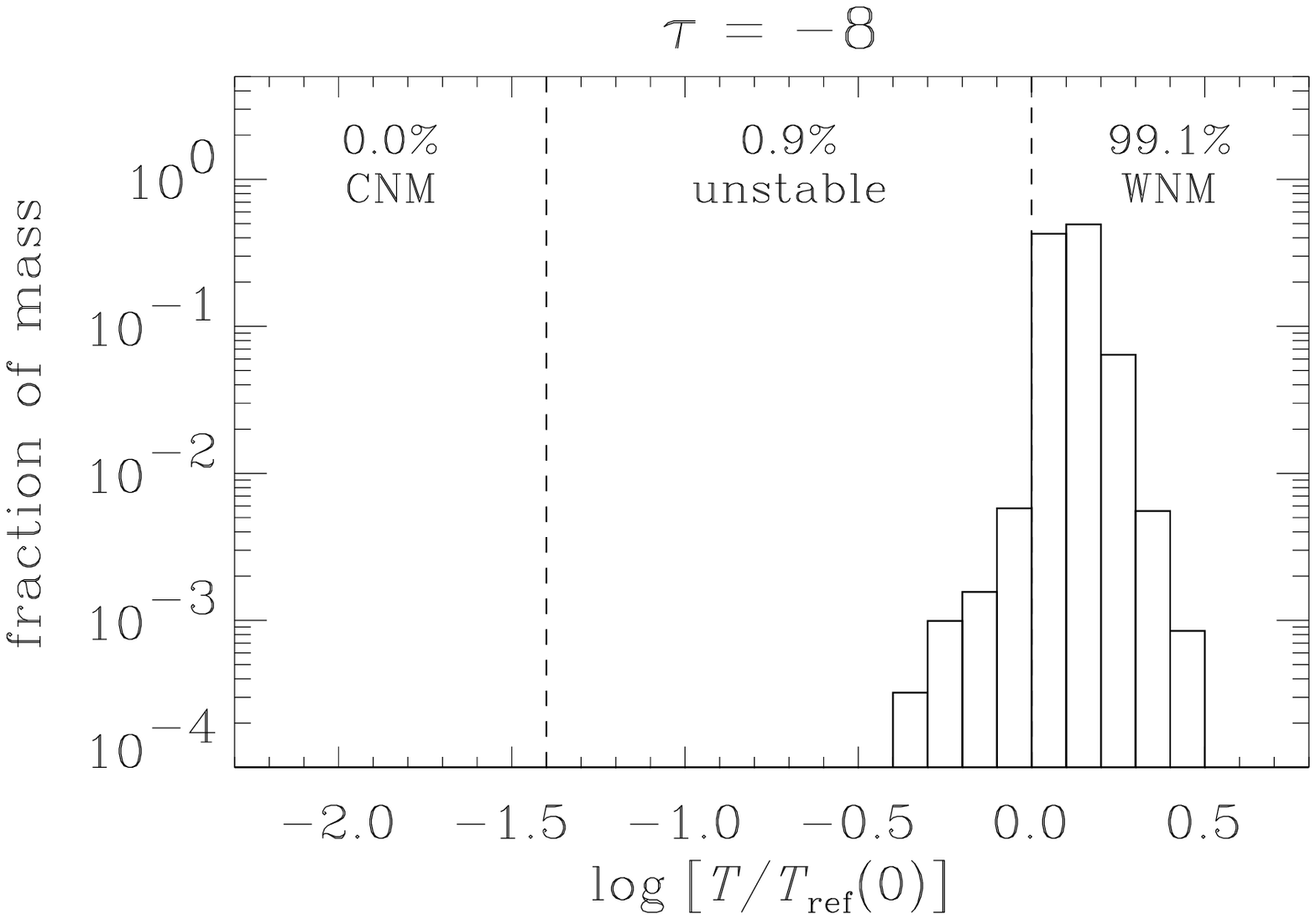}
\newline
\newline
\newline
\includegraphics[width=3.2in]{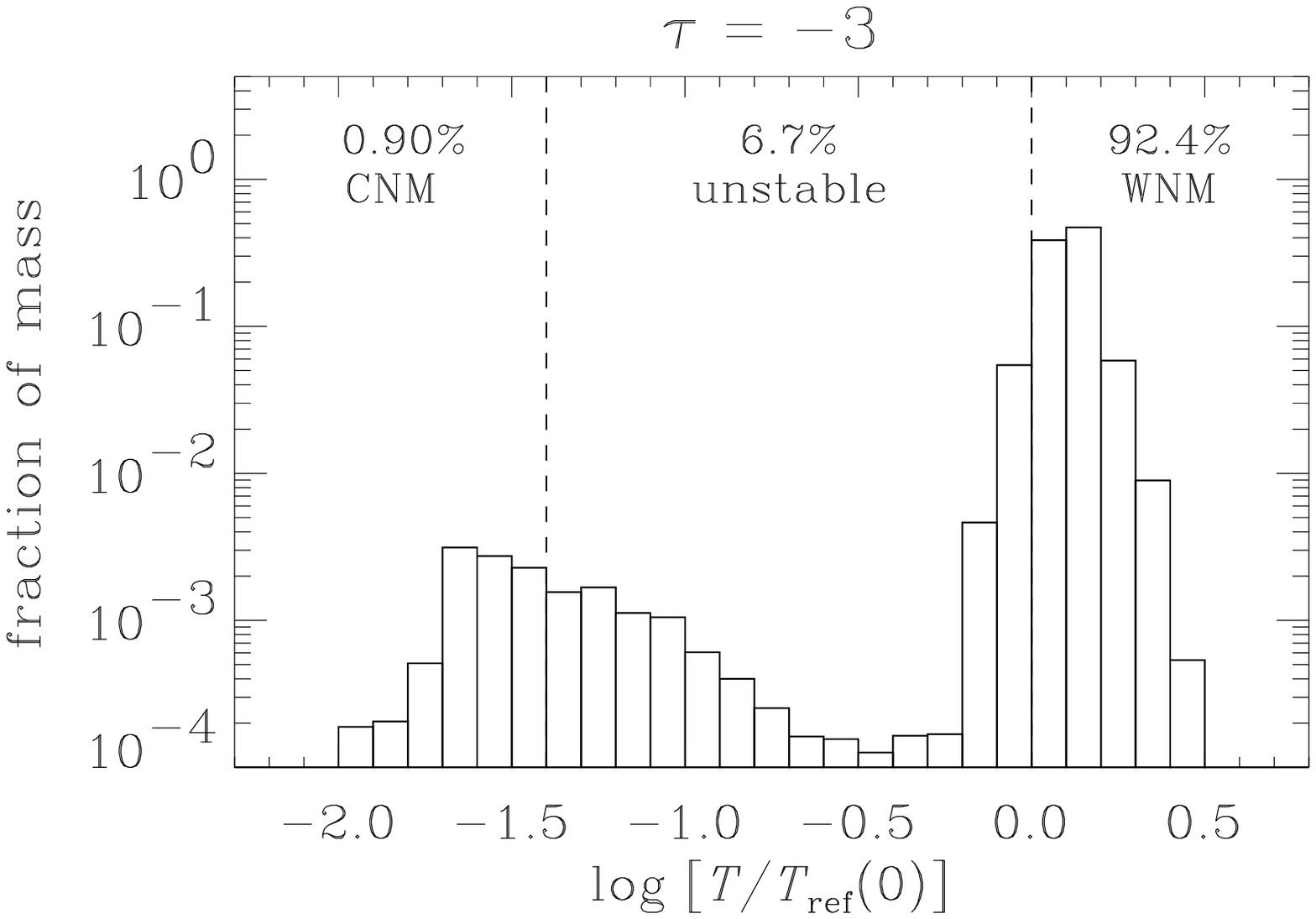}
\qquad\quad
\includegraphics[width=3.2in]{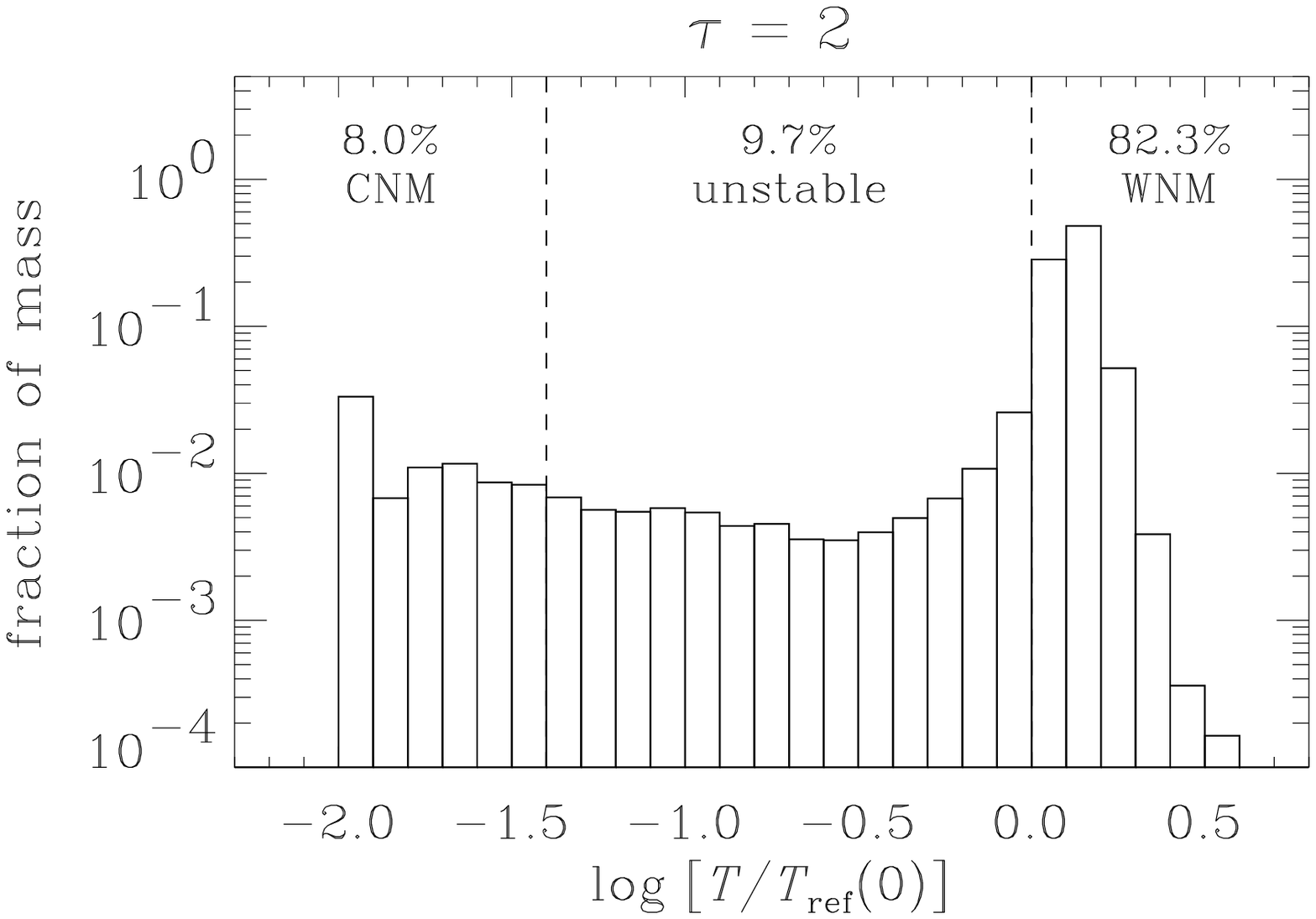}
\newline
\newline
\newline
\includegraphics[width=3.2in]{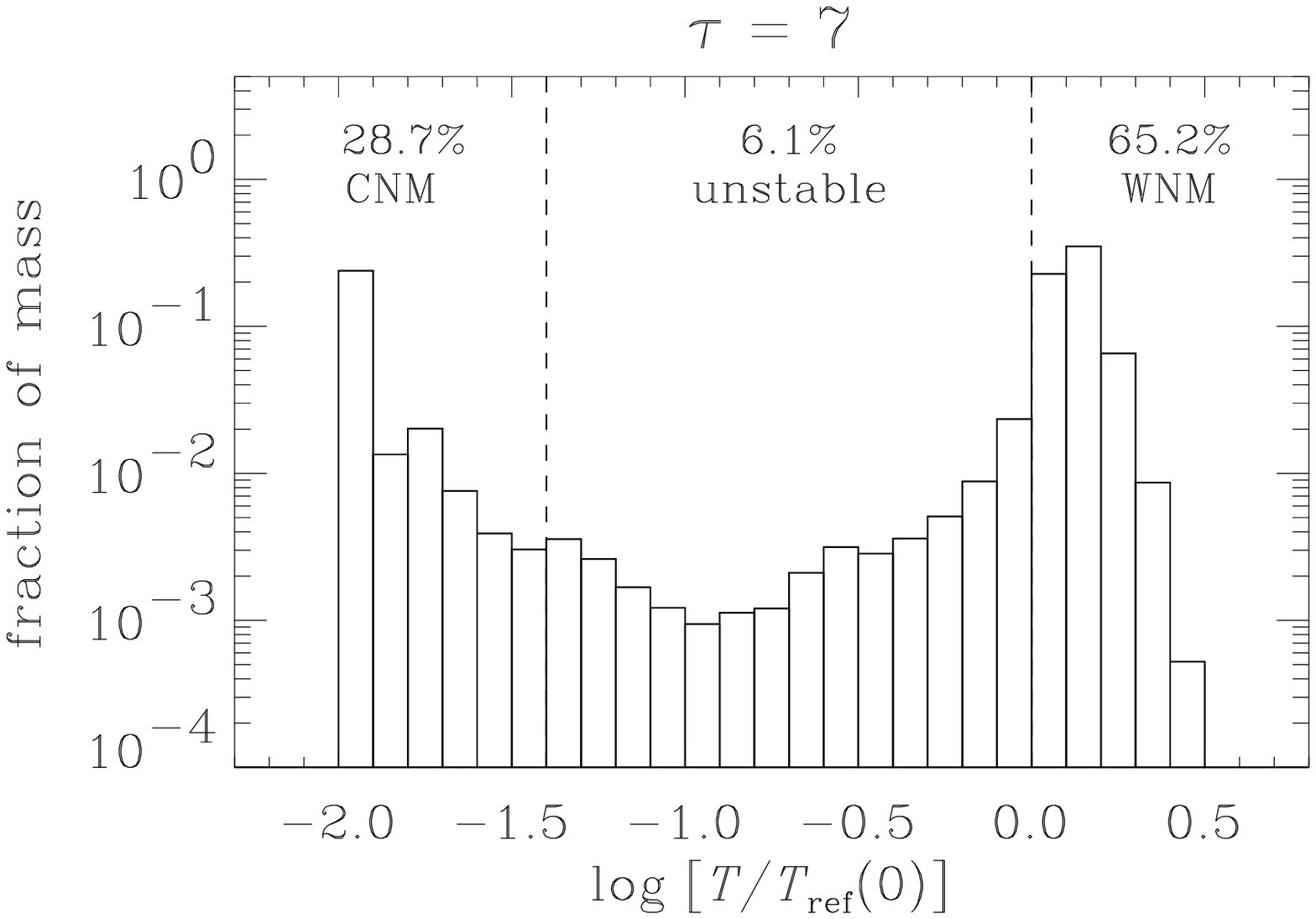}
\qquad\quad
\includegraphics[width=3.2in]{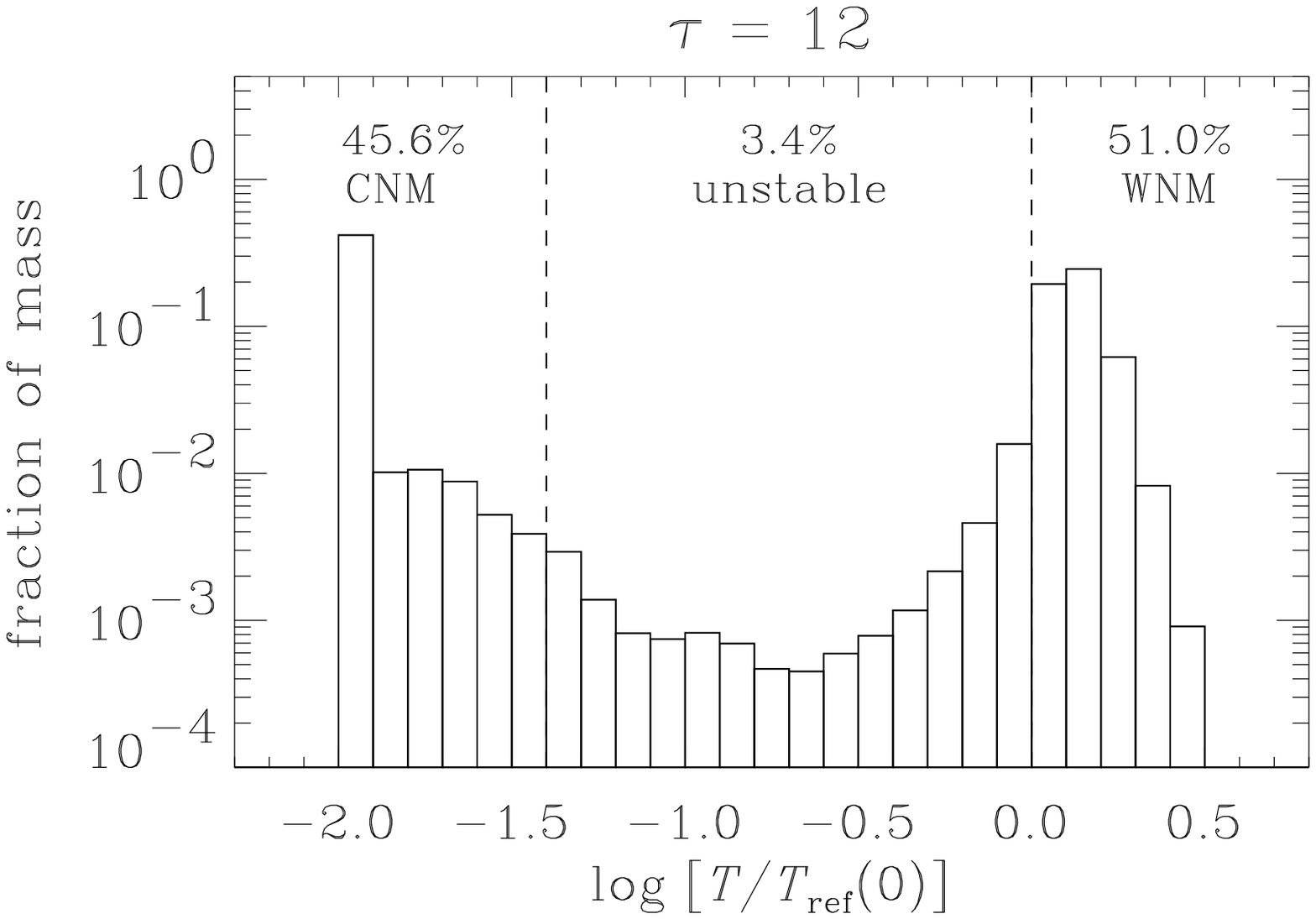}
\newline
\caption{Fraction of the total mass in each thermal phase (CNM,
unstable, and WNM) at times $\tau = -13$, $-8$, $-3$, 2, 7, and
12. Mass is binned by temperature in intervals of width
$\Delta\log[T/T_{\rm ref}(0)]=0.1$. The vertical dashed lines
denote the temperature boundaries separating the different thermal
phases. The total mass fraction in each of the phases is given as
a percentage. The unit of time is $C_{\rm ref}(0)/g =$ 6.05 Myr.}
\label{F_tdist}
\end{figure*}

In light of the recent H\textsc{i} surveys by \citet{ht03}, it is
of interest to calculate the fractions of interstellar mass in
each thermal phase. In Figure \ref{F_tdist}, we show histograms of
the fraction of the total mass in each thermal phase (CNM,
unstable, and WNM) at times $\tau=-13$, $-8$, $-3$, 2, 7, and 12.
Mass is binned by temperature in intervals of width
$\Delta\log[T/T_{\rm ref}(0)]=0.1$. The vertical dashed lines
denote the temperature boundaries separating the different thermal
phases ($\widetilde{T}_{\rm min}$, $\widetilde{T}_{\rm max}$). The
mass fraction in each phase is shown in each figure as a
percentage. The reference state ($\tau=-13$) has all the mass
in the WNM. As the Parker instability progresses, phase
transitions are triggered, leading to a transfer of mass from the
WNM, through the thermally-unstable phase, into the CNM. At the
time when cold H\textsc{I} clouds have been formed and have
settled into their final quasi-equilibrium states, the percentages
of mass in the CNM and WNM are approximately 40\% and 60\%,
respectively. (This number is found by simply averaging the
results between $\tau=7$ and 12.) Enough time has elapsed for
there to be an insignificant amount of mass in the
thermally-unstable phase.

Taken at face value, this calculation predicts the existence of
cold H$\textsc{i}$ clouds with maximum number density $\simeq
10^3$ cm$^{-3}$, maximum magnetic field strength $\simeq
20\,\mu$G, and temperatures $\simeq 50$ K. The {\em mean} values
of the density and field strength in the cold clouds are found to
be $\simeq 20$ cm$^{-3}$ and $\simeq 4.3$ $\mu$G, respectively.
The rms value of the magnetic field strength in the CNM is
$\simeq 6.4$ $\mu$G, while the maximum line-of-sight value of the
field occurs for a line-of-sight in the azimuthal direction in the
galactic plane and is $\simeq 13.5$ $\mu$G. The (two-dimensional)
turbulent velocity dispersion inside the CNM clouds is $\simeq 1.6$
km s$^{-1}$; this corresponds to a (thermal) Mach number $\simeq 2.8$.
The mass of a typical cold cloud is calculated by assuming a thickness
in the third (radial) direction corresponding to that of a galactic
shock ($\simeq 100$ pc); we thus find the typical mass to be $\simeq
10^5$ M$_\odot$. The spatial separation between cloud complexes is
$\simeq 500$ pc. An estimate of the maximum mass-to-flux ratio of
typical cold H$\textsc{i}$ clouds yields
$(M/\Phi)/(M/\Phi)_{\rm crit} \simeq 0.1 - 0.3$, where
$(M/\Phi)_{\rm crit}\equiv (63G)^{-1/2}$ is the critical
mass-to-flux ratio calculated by \citet{ms76}.

\section{Summary and Discussion}\label{S_summary}

We have performed two-dimensional simulations of the nonlinear
evolution of the Parker instability properly accounting for phase
transitions between the warm and cold neutral media. The evolution
is initiated by small-amplitude velocity perturbations that have,
and at later times lead to, evolution that preserves the symmetry
of the dominant midplane-crossing (`even') mode --- the mode that
allows field lines to cross the galactic plane. This mode has been
shown by BMP97 to be naturally selected when starting from a
spectrum of random velocity perturbations. The {\it qualitative}
features of the nonlinear evolution are similar to those found by
calculations that ignored phase transitions: arches of magnetic
field lines rise high above the galactic plane, accompanied by
accumulations of mass in the resulting magnetic valleys. However,
there are important {\it quantitative} differences between the two
sets of calculations. Phase transitions, a consequence of
realistic heating and cooling included in the present calculation,
lead to structures (cold clouds) with typical densities much
greater and temperatures much smaller than those achieved by
isothermal calculations, and which are in excellent quantitative
agreement with observations of cold H\textsc{i} clouds.

The main result of this work is that the Parker instability is
{\em directly} responsible for the formation of interstellar
clouds and cloud complexes, despite recent claims to the contrary
by Kim et al. (1998, 2001, 2002) and \citet{skfmhr00}. Cold
H\textsc{i} clouds are seen to form whose masses ($\simeq 10^5$
M$_\odot$), mean magnetic field strengths ($\simeq 4.3\,\mu$G),
rms magnetic field strengths ($\simeq 6.4\,\mu$G), mean densities
($\simeq 20$ cm$^{-3}$), mass-to-flux ratios ($\simeq
0.1-0.3$ relative to critical), temperatures ($\simeq 50$ K),
turbulent velocity dispersions ($\simeq 1.6$ km s$^{-1}$), and
spatial separations ($\simeq 500$ pc) are all in agreement with
observations. The maximum number density is $\simeq 10^3$
cm$^{-3}$ and the maximum magnetic field strength $\simeq
20\,\mu$G. These physical conditions are such that, had we
included the additional phase transition from atomic to molecular
hydrogen, molecular clouds would have formed very rapidly within
these giant H\textsc{i} clouds.

We find that the fraction of ISM mass in the WNM is approximately
60\%, in excellent agreement with observations by \citet{ht03}.
The bulk of the CNM has a sheet-like morphology. We also note the
presence of cold `cloudlets' of size $\simeq 1$ pc strung along
field lines. These cloudlets are both qualitatively and
quantitatively similar to those first discovered by
\citet{heiles67}. The median column density of a typical cloudlet
is $\simeq 0.3\times 10^{20}$ cm$^{-2}$.

An apparent discrepancy between this work and the observational
results of \citet{ht03} is the amount of WNM in the
thermally-unstable phase. In the simulation, the maximum
percentage of WNM in the unstable phase is $\simeq 10\%$. By
contrast, the observed percentage is $\simeq 48\%$ by
\citet{ht03}. The discrepancy is easily understood as a
consequence of the fact that the observations see the ISM as it is
{\it today}, after clouds not only have formed but also some of
them have given birth to stars, including OB associations, which
alter the physical conditions of their parent clouds as well as
the amount of matter in the warm unstable phase. In other words,
to account for the observed amount of gas in the thermally
unstable phase, a calculation must include feedback from at least
some of the energetic events that occur on relatively short
timescales after clouds and stars have formed (e.g., stellar
winds, UV radiation, or supernovae). Including such effects is
beyond the scope of the present study, whose focus is cloud
formation, not events that follow star formation.

Our results differ substantially from those of the recent
three-dimensional simulations by \citet{kh07}. Important
differences also exist in the formulation of the problem. Their
neglect of thermal conductivity, together with their very low
resolution ($60\times 260\times 100$ compared to our $2048\times
4096$), results in thermally-unstable modes showing unresolved
growth on their grid size of 10 pc (our grid resolution is $\simeq
0.2$ pc). In addition, their assumed symmetry about the galactic
plane limits the problem to the `odd' mode, which is not favored
during the nonlinear evolution over the `even' mode we have
followed. Their heating and cooling functions imply a stable CNM
phase at an unrealistically-high temperature of $\simeq 3000 -
4000$ K (depending on the model). For the realistic heating and
cooling functions we have employed, these temperatures lie in the
thermally-unstable regime; we find that the thermally stable CNM
phase exists below a temperature of 185 K.

The results of this work lend strong support and quantify the
scenario of cloud formation envisioned early on by Mouschovias et
al. (1974). Taken in conjunction with the many successes of the
theory of magnetic support of self-gravitating clouds and the
ambipolar-diffusion--initiated fragmentation and star formation in
such clouds (e.g., see \S~4.2 of Mouschovias et al. 2006), the
present work shows that magnetic fields play a crucial role not
only in the formation of stars, but also in the formation and
early evolution of their birthplaces, the interstellar clouds.

\section*{Acknowledgments}

We thank Shantanu Basu for providing the code developed for BMP97,
which was used in an early phase of this work, as well as Carl
Heiles and Konstantinos Tassis for useful discussions. TM
acknowledges partial support from the National Science Foundation
under grant NSF ASF-07-09206. All computations were performed on
the {\it Turing} cluster (a 1536-processor Apple G5 X-serve
cluster devoted to high-performance computing in engineering and
science) maintained and operated by the Computational Science and
Engineering Program of the University of Illinois at
Urbana-Champaign.

\label{lastpage}

\end{document}